\documentclass[acmlarge]{acmart}

\usepackage{amsmath}
\usepackage{amsfonts}
\usepackage{graphicx}
\usepackage{textcomp}
\usepackage{xcolor}
\usepackage{bm}
\usepackage{setspace}
\usepackage{enumitem}
\usepackage{makecell}
\usepackage{url}
\usepackage{multirow}
\usepackage{epstopdf}
\usepackage{color}
\usepackage{colortbl}
\usepackage{hyperref}
\usepackage[ruled]{algorithm2e}
\usepackage[small]{subfigure}

\SetKwRepeat{Do}{do}{while}
\SetKwComment{Comment}{/* }{ */}

\newcommand{\n}{GUFU}

\AtBeginDocument{%
  }

\setcopyright{acmlicensed}
\copyrightyear{2025}
\acmYear{2025}
\acmDOI{10.1145/3712277}

\acmJournal{IMWUT}
\acmVolume{9}
\acmNumber{1}
\acmArticle{3}
\acmMonth{3}

\begin{document}

\title{Graph-based Fingerprint Update Using Unlabelled WiFi Signals}

\author{Ka Ho Chiu}
\email{khchiuac@connect.ust.hk}
\orcid{0000-0002-4480-1868}
\affiliation{
  \institution{The Hong Kong University of Science and Technology}
  \city{Hong Kong}
  \state{Hong Kong}
  \country{China}
}

\author{Handi Yin}
\email{hyin335@connect.hkust-gz.edu.cn}
\orcid{0009-0004-7287-3879}
\affiliation{%
  \institution{The Hong Kong University of Science and Technology (Guangzhou)}
  \city{Guangzhou}
  \state{Guangdong}
  \country{China}
}

\author{Weipeng Zhuo}
\email{weipengzhuo@uic.edu.cn}
\authornote{Corresponding author.}
\orcid{0000-0002-1810-7071}
\affiliation{%
  \institution{Guangdong Provincial/Zhuhai Key Laboratory of IRADS, and \\ Department of Computer Science, BNU-HKBU United International College}
  \city{Zhuhai}
  \state{Guangdong}
  \country{China}
}

\author{Chul-Ho Lee}
\email{chulho.lee@txstate.edu}
\orcid{0000-0002-4778-8996}
\affiliation{
  \institution{Texas State University}
  \city{San Marcos}
  \state{Texas}
  \country{United States}
}

\author{S.-H. Gary Chan}
\email{gchan@ust.hk}
\orcid{0000-0003-4207-764X}
\affiliation{
  \institution{The Hong Kong University of Science and Technology}
  \city{Hong Kong}
  \state{Hong Kong}
  \country{China}
}

\renewcommand{\shortauthors}{Chiu et al.}

\begin{abstract}
WiFi received signal strength (RSS) environment evolves over time due to the movement of access points (APs), AP power adjustment, installation and removal of APs, etc.  We study how to effectively update an existing database of fingerprints, defined as the RSS values of APs at designated locations, using a batch of newly collected unlabelled (possibly crowdsourced) WiFi signals.  Prior art either estimates the locations of the new signals without updating the existing fingerprints or filters out the new APs without sufficiently embracing their features.  
To address that, we propose \n{}, a novel effective {\bf g}raph-based approach to {\bf u}pdate WiFi {\bf f}ingerprints using {\bf u}nlabelled signals with possibly new APs.
Based on the observation that similar signal vectors likely imply physical proximity, \n{} employs a graph neural network (GNN) and a link prediction algorithm to retrain an incremental network given the new signals and APs.  After the retraining, it then updates the signal vectors at the designated locations.
Through extensive experiments in four large representative sites, \n{} is shown to achieve remarkably higher fingerprint adaptivity as compared with other state-of-the-art approaches, with error reduction of 21.4\% and 29.8\% in RSS values and location prediction, respectively.
\end{abstract}





% \received{1 May 2024}
% \received[revised]{1 November 2024}
% \received[accepted]{31 December 2024}

\maketitle

\section{INTRODUCTION}
\label{sec:intro}
A WiFi fingerprint is defined as the received signal strength (RSS) values of the WiFi access points (APs) at a location. With the proliferation of APs and WiFi-enabled mobile devices~\cite{zhang2019deep,zhu2020indoor,liu2022survey, shang2022overview}, WiFi fingerprints have found extensive applications in indoor navigation~\cite{dong2022lstmnavigation, lin2020locater, fan2021siabr, guo2022wepos, rihan2022hybridfingerprint}, WiFi service analysis~\cite{determe2022monitoring, sarcevis2023fusionrheatmap, marakkalage2021urbanheatmap, turetta2022csi}, geofencing~\cite{zhuo2023geofencing, lin2021tcove, lu2019noisygeofencing, guo2021indoormonitor, hegde2023coverage}, healthcare~\cite{trivedi2021trace, ge2022contactless, zheng2023health}, etc.

\begin{figure}
    \centering
    \subfigure[Before.]{
    \includegraphics[width=0.4\linewidth]{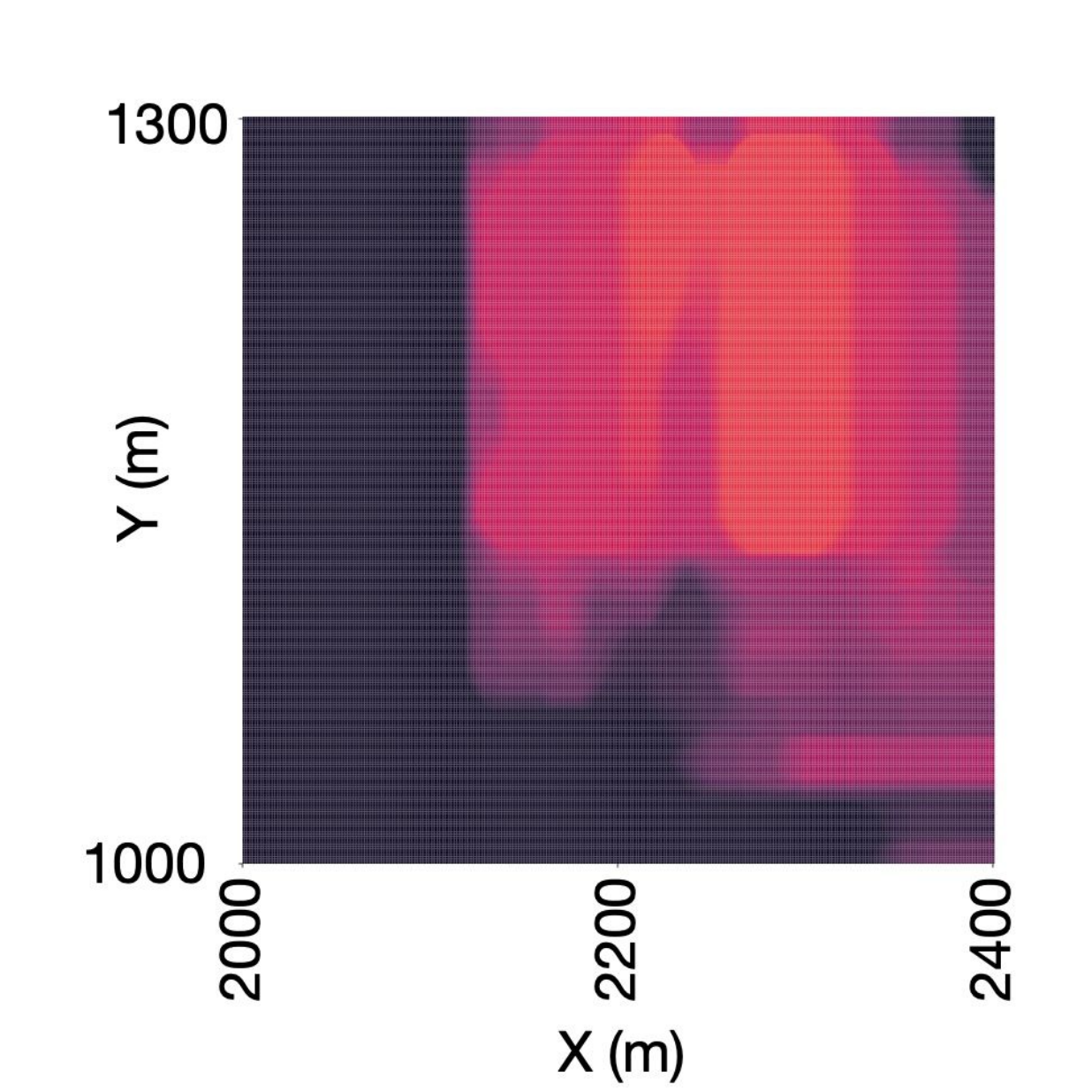}
    }  
    \subfigure[After.]{
    \includegraphics[width=0.4\linewidth]{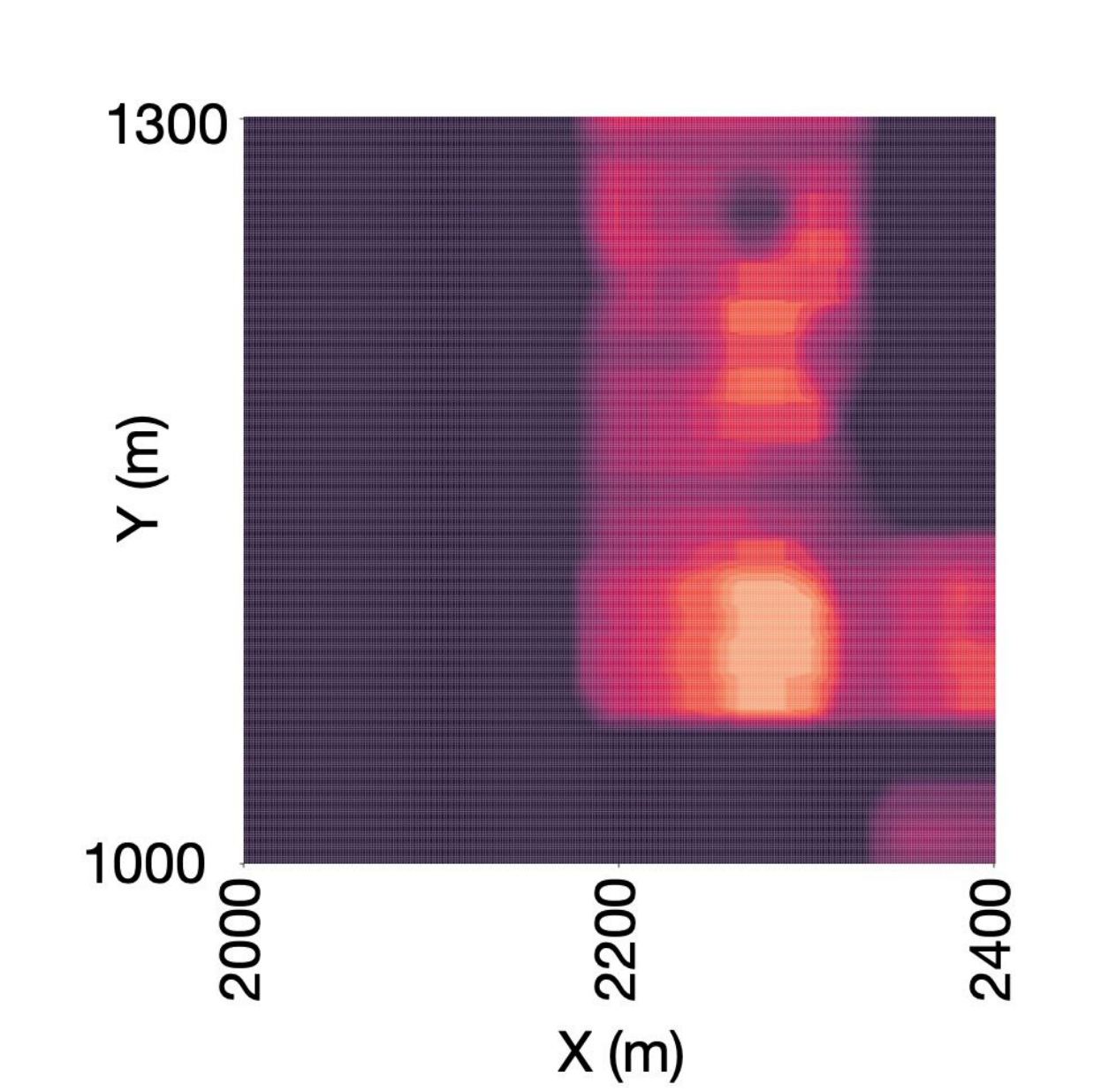}
    }
    \vspace{0.05in}
    \subfigure
    {\includegraphics[width=0.11\linewidth]{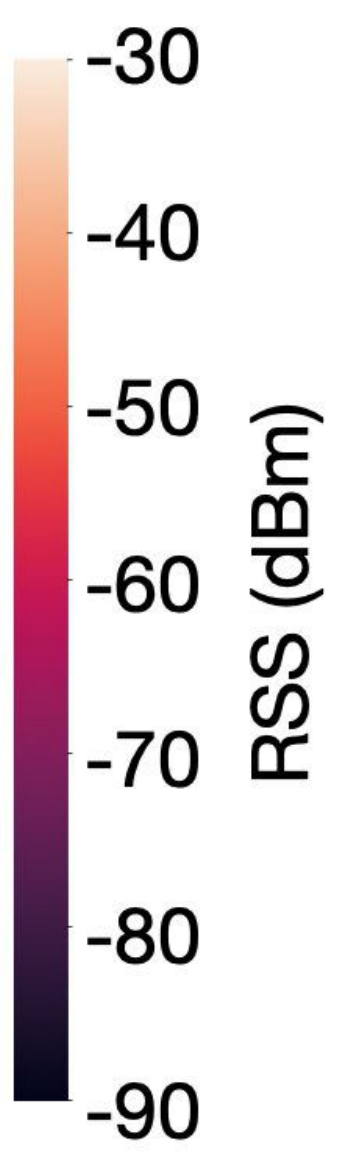}} 
    \caption{Comparison of signal distributions from AP with SSID 34F8E7C0C583 over two months in a campus.}
    \label{fig:intro}
    \Description[AP change]{Comparison of signal distributions from AP with SSID 34F8E7C0C583 over two months in a campus.}
\end{figure}

A Wifi fingerprint database is often initially obtained with a full site survey where WiFi signals, given by AP IDs and their RSSs, are collected at designated locations (for example, over a grid size of a few meters). As WiFi environments evolve due to factors such as AP power adjustment, re-location, installation and removal, the existing fingerprint has to be updated over time to maintain service quality~\cite{he2017laafu,gao2020crisloc,xiang2021self,tian2021domainupdate, li2021imgadversarial, wang2023mtdan}. Table~\ref{tab:ap} shows a WiFi environment on a campus that undergoes renovations, showcasing that the number of detected APs from multiple floors can significantly change over a two-month period. While some APs remain constant, numerous APs are removed, and a significant number of new APs are discovered. Figure~\ref{fig:intro} demonstrates an RSS heatmap of an AP that remains in the same area throughout this period, highlighting the notable differences in the spatial distribution of signals. These differences indicate that the AP power and/or location may have been changed.

In order to update an existing fingerprint database, traditionally new blanket site surveys are conducted regularly. As such surveys are labor-intensive and time-consuming, crowdsourcing approaches have been proposed, where new signals are collected spatially and randomly over time by mobile user devices in the venue {\em without} any location labels. These new unlabeled signal samples, sampled effortlessly at different locations, 
are processed periodically in batches, typically once every week or so, to update the existing fingerprints (RSS at existing designated locations).   

\begin{table}
\caption{Number of APs over a two-month period in a campus.}
\label{tab:ap}
\begin{center}
\begin{tabular}{| c | c | c |} 
 \hline
Before & After & Overlapped/Shared APs\\ \hline
 \hline
 494 & 517 & 193 \\
 \hline
\end{tabular}
\end{center}
\vspace{-0.2in}
\end{table}

In this work, we study the challenging problem of how to effectively update an existing fingerprint database given a batch of unlabelled (crowdsourced) signals that may consist of new APs. Our goal is to make full use of the signal features, including those of the new APs, to accurately update the signal vectors at the designated locations of the existing fingerprints.  

Previous works on fingerprint update often estimate the locations of the newly collected signal samples and add them as new fingerprints into the existing database~\cite{li2020pseudolabels, song2019cnnloc,abbas2019wideep, zhang2023dagraph}. While the new signals are included, the existing, possibly outdated, fingerprints, however, remain \emph{unchanged} in the database. These outdated fingerprints can contaminate the database rather quickly over time. To tackle that, some recent approaches update the existing fingerprints by merging the new signals into the existing ones~\cite{ganin2015dann,li2021imgadversarial,chen2020csi,chen2022fidora, wang2023mtdan}. These methods, however, consider only the common APs without the \emph{new} APs in the crowdsourced samples. This adversely affects the update process, because the information or features of the new APs have not been embraced or exploited. 

We propose \n{}, a novel effective {\bf g}raph-based approach to {\bf u}pdate WiFi {\bf f}ingerprints using {\bf u}nlabelled signals with possibly new APs. To the best of our knowledge, this is the first work on how to effectively update the existing (aged) fingerprints with a batch of newly collected unlabeled signals by considering all APs, including the shared and new ones.

\n{} consists of offline and online stages. In the offline stage, \n{} is initialized with a bootstrap site survey, where fingerprints are collected at designated grid locations, typically a few meters apart. Using the fingerprints, \n{} trains its two components, namely, a feature extractor and a graph neural network (GNN). The feature extractor extracts signal features from RSSs. The extracted features are then used to create a weighted graph formed by AP nodes and sample nodes, in which those features are fed as the initial sample node features. Subsequently, edges are created between the two types of nodes based on the signal strengths, while AP node features are initialized as the weighted average of the features from neighboring sample nodes. Additionally, observing that similar signal features likely indicate location proximity, extra virtual edges between sample nodes are added to the graph. The GNN is finally trained on this weighted graph, utilizing both node features and edge features, which are determined through weighted feature aggregation between the nodes.

After the offline stage, \n{} then enters into the online stage where the fingerprints evolve with the WiFi environment based on batches of unlabeled (crowdsourced) signals collected over time. New sample nodes for the graph are created using the shared APs between the existing fingerprints and each new batch of signals, along with nodes for any new APs that appear only in the new batch. The trained feature extractor and GNN are then applied to get those new nodes' features. These features are subsequently used in two MLPs to update the existing node features. The updated features are finally used to amend the corresponding signal strengths in the existing fingerprints.

To effectively address the evolving features of both new and outdated APs, \n{} employs a novel edge prediction algorithm to update the edges in the graph. This algorithm establishes new edges between the new APs and the labeled samples while removing potentially outdated edges. With these updated edges, the GNN can be updated to have its node features better represent the signal dynamics. Then by utilizing the trained feature extraction and node feature aggregation, the features of the new APs can also be leveraged to update the RSS values of the existing fingerprints.

Our contributions are summarized as follows:
\begin{itemize}

\item We propose \n{}, a novel graph-based approach to update WiFi fingerprints using unlabeled (crowdsourced) signals with possibly new APs. \n{} effectively amends each of the RSS values in the existing fingerprints at designated locations by embracing the features of the new APs.

\item We propose a novel edge prediction module in \n{} to incorporate the features of the new APs. This module establishes connections between the new APs and existing samples in GNN to greatly improve the effectiveness of fingerprint updates. In addition to embracing new APs, this module can also remove existing yet outdated APs and their associated edges via a forgetting mechanism.

\item We conduct extensive experimental studies on \n{} in four different major sites, namely, a campus building, a hospital and two shopping malls, over a long period of time (eight months). Our experimental results demonstrate that the fingerprints are effectively updated over time as the WiFi environment evolves. \n{} outperforms state-of-the-art algorithms significantly, reducing 21.4\% in RSS error and 29.8\% in signal location error.

\end{itemize}

The rest of this paper is organized as follows. After reviewing the related work in Section~\ref{sec:related}, we overview  \n{} in Section~\ref{sec:sys_overview}. In Section~\ref{sec:init} we discuss the offline training of \n{} given an initial bootstrapped fingerprints by a site survey.  In Section~\ref{sec:update} we present the batch update of fingerprints for existing APs, while in Section~\ref{sec:update2} We present the fingerprint update with AP changes, i.e., addition and removal of APs. We discuss experimental results in Section~\ref{sec:eva} and conclude in
Section~\ref{sec:conclude}.

\section{RELATED WORK}
\label{sec:related}
\noindent \textbf{Database update by appending new fingerprints:} Many previous approaches predict locations for new signals and add them to the existing database. For the prediction task, earlier works~\cite{huang2019gprpdr,xu2015imgupdate,wei2021mm} leverage additional information from IMU sensors or cameras in addition to WiFi signals. Other recent studies~\cite{song2019cnnloc,abbas2019wideep,chen2020csi,huang2020wifi,yang2021robust, dang2023cluster, rajab2023fuse, yang2022indoor, zhang2023dagraph, chen2024dafingerprint} focus on WiFi signals and aim to train a WiFi-based classifier to find the most similar fingerprint records in the database for each new signal. For example, CNNLoc~\cite{song2019cnnloc} combines a one-dimensional CNN with a stacked autoencoder for classifying WiFi signals. WiDeep \cite{abbas2019wideep} utilizes a similar autoencoder model and combines it with a noise injector to better extract AP-invariant features. FIDo~\cite{chen2020csi} uses WiFi channel state information and builds a variational autoencoder to classify WiFi signals. WiDAGCN~\cite{zhang2023dagraph} models APs and signal samples into a graph and utilizes graph attention to model similarities between existing and new signals. These approaches add new signals with their predicted locations into the database. However, the existing and possibly outdated records in the database remain \emph{unchanged}, which deteriorates the quality of fingerprints over time. By contrast, \n{} updates the existing (aged) fingerprints by estimating the RSS values that have been possibly outdated for each designated location in the database so that every record in the database can be up to date.

\vspace{1mm}
\noindent \textbf{Database update by fusing new signals with existing fingerprints:} Aged signals for the existing fingerprints and newly crowdsourced ones obtained from close locations may follow similar distributions. Hence, recent approaches~\cite{tian2021domainupdate,li2021imgadversarial,chen2020csi,chen2022fidora, li2022NQRELoc, zhang2023adaptivefewshot, wang2023mtdan, chen2024update} attempt to fuse similar signals collected at \emph{different} times to update fingerprints. For example, TransLoc~\cite{tian2021domainupdate} and iToLoc~\cite{li2021imgadversarial} aim to extract temporal-invariant features to train a classification model such that similar signals can be merged. In addition, Fidora~\cite{chen2022fidora} reconstructs old and new signals together via semi-supervised learning and updates signals in the old fingerprints. Another recent work MTDAN~\cite{wang2023mtdan} further utilizes multi-target domain adaptation for signal updates. These approaches, however, only consider the shared APs that appear in the existing fingerprints. As a result, they have not fully exploited the signal features of the new APs, which could take a significant share in the newly collected signals, as shown in Table~\ref{tab:ap}. By contrast, we propose an edge prediction module in \n{} to associate the new APs with the existing fingerprints.  This allows us to leverage the features from the new APs in the process of updating the aged fingerprints.

\vspace{1mm}
\noindent \textbf{Graph-based radio frequency (RF) signal modeling:} RF signals such as WiFi signals are traditionally processed in the form of fixed-length vectors or matrices. This approach, however, may suffer from the missing-value problem~\cite{zhuo2022grafics}. Specifically, not all the signals from the APs at a location may be fully scanned, thereby leaving several entries in the vectors or matrices empty. These missing entries are usually filled with arbitrarily small values, but they may introduce unintended artifacts in the feature learning process. To solve this problem, several recent studies~\cite{cherian2018graph, zhuo2022grafics, zhang2020graph, yang2022gat, zhang2023dagraph,  zhuo2023fisoneflooridentificationlabel} utilize a graph to model the RF signals. Zhang et al.~\cite{zhang2020graph} propose to use a homogeneous graph to model the relationships among signal samples, while WiDAGCN~\cite{zhang2023dagraph} models APs and signal samples separately. Besides utilizing a bipartite graph model with APs (given by MAC addresses) and signal samples being two different types of nodes, \n{} creates virtual edges to better capture their similarity. With this novel design, \n{} is able to effectively propagate possible changes in RSS values via edges in the graph and in turn better update the fingerprints.

\section{SYSTEM OVERVIEW}
\label{sec:sys_overview}
\begin{figure*}
	\centering
	\includegraphics[width=\textwidth]{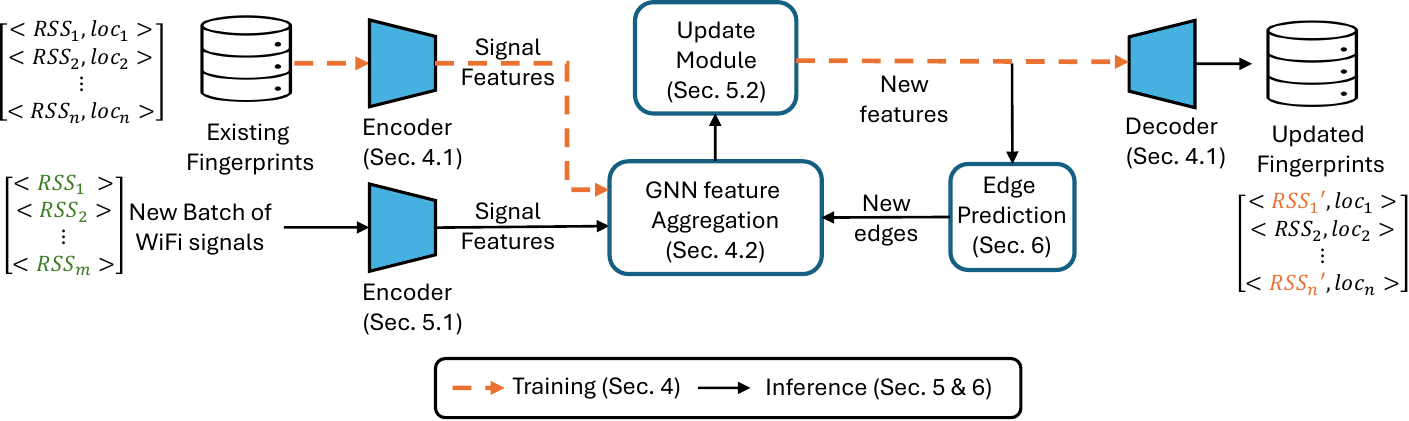}
	\caption{A system overview of \n{}.}
	\label{fig:sys_overview}
    \Description[System Overview]{Showing the general structure of \n{}, especially for the one involved in training.}
\end{figure*}

We overview \n{} in Figure~\ref{fig:sys_overview}. \n{} is bootstrapped with an existing fingerprint database, obtained through surveys at designated locations. This database is used for initialization, in which the RSS feature extractor is trained (Section~\ref{subsec:RSS_extractor}) and the GNN is constructed using the signal features obtained by the trained feature extractor (Section~\ref{subsec:graph_training}). After that, given a newly collected batch of unlabeled signals, \n{} checks whether it has new APs apart from those in the existing fingerprints. For those existing APs, their signal features are obtained via inference through the GNN (Section~\ref{subsec:graph_inference}), and new corresponding signal nodes are added to the GNN to infer the locations of new samples and the updated feature vectors for the signals (Section~\ref{subsec:loc_predictor}). For the new APs, an additional edge predictor is trained (Section~\ref{subsec:new_ap_pred}) to create edges between those new APs and existing samples in the GNN. At the same time, a forgetting mechanism is applied to remove edges connected to potentially outdated APs as well as the nodes that correspond to the outdated APs~(Section~\ref{subsec:ap_removal}). These changes in AP nodes and edges enable the feature extractor and the GNN to be updated, and the updated feature vectors can be obtained from the updated GNN. Finally, those updated vectors are fed to the decoder for outputting the updated signal strengths in the existing fingerprint. To help better understand the operations of \n{}, in Table~\ref{tab:notations}, we collect all the key notations used throughout the paper.

\begin{table}
\caption{Summary of notations.}
\centering 
\begin{tabular}{r p{10cm} }
\toprule
$\bm{X}$ & Set of RSS values in the fingerprint database \\
$\bm{Y}$ & Set of locations in the fingerprint database \\
$\bm{U}$ & Set of RSS values in a new batch of signals \\
$\bm{Z_X}$ & Set of signal features for $\bm{X}$\\
$\bm{Z_U}$ & Set of signal features for $\bm{U}$\\
$\bm{V}$ & Set of nodes in graph $\mathcal{G}$ \\
$\bm{v}_x$ & A sample node in $\bm{V}$, with $x\in \bm{X}$\\
$\bm{v}_m$ & An AP node in $\bm{V}$\\
$\bm{z_x}$ & Node feature for node $\bm{v}_x$\\
$\bm{z_m}$ & Node feature for node $\bm{v}_m$\\
$\bm{E}$ & Set of edges in graph $\mathcal{G}$ \\
$\bm{E}_{virtual}$ & Set of virtual edges created in graph $\mathcal{G}$ \\
$\bm{z_{xm}}$ & Edge feature for edge $\bm{e}_{xm} \in \bm{E} \cup \bm{E}_{virtual}$\\
$\bm{W}$ & Set of edge weights in graph $\mathcal{G}$ \\
$\mathbb{W}_0$, $\mathbb{W}_1$, $\mathbb{W}_2$ & Trainable weights in the GNN \\
$\bm{G(v)}$ & Goodness score for a node $v$ in graph $\mathcal{G}$ \\
$\bm{F(v)}$ & Fairness score for a node $v$ in graph $\mathcal{G}$ \\
\bottomrule
\end{tabular}
\vspace{0.1in}
\label{tab:notations}
\end{table}

\section{TRAINING WITH BOOTSTRAPPED FINGERPRINTS}
\label{sec:init}
In this section, we introduce the initialization of \n{} during the offline stage. \n{} is bootstrapped using a set of fingerprints collected through a site survey. These fingerprints serve as the input for training the RSS feature extractor (Section~\ref{subsec:RSS_extractor}). After training, the extractor can generate the embeddings for any fingerprints. The embeddings of the existing fingerprints are then used to train a GNN on a weighted graph, along with additional virtual edges (Section~\ref{subsec:graph_training}).

\subsection{Encoder and Decoder for Feature Extraction}
\label{subsec:RSS_extractor}
The existing WiFi fingerprint database consists of signals and their corresponding location labels, typically obtained through surveys at predetermined locations. Each signal sample is obtained from a WiFi scanning and contains pairs of detected MAC addresses (MACs) along with their associated received signal strength (RSS) values. 
Let $\bm{Y}\in \mathbb{R}^{N\times 2}$ represent the set of locations in the fingerprint database, where $N$ is the number of samples with location labels. Additionally, let $\bm{X} \in \mathbb{R}^{N\times n_s}$ denote the corresponding set of RSS values, where $n_s$ is the total number of MACs detected at each location. If a particular MAC is not detected in a sample, we assign its corresponding entry in $\bm{X}$ a value of $-120$ dBm, which is the common practice in recent studies~\cite{zhuo2022grafics, zhang2020graph, zhang2023dagraph, zhuo2023fisoneflooridentificationlabel}. To normalize the RSS values to $\left[0, 1\right]$, we add an offset of $120$ dBm to each entry in $\bm{X}$ and divide the resulting value by 120. We refer to $\bm{X}$ as the set of \emph{normalized} RSS values for the rest of this paper.

In addition to the initial fingerprint database, RSS values are continuously collected in a crowdsourced manner, without accompanying location labels. A new batch of unlabeled samples becomes available at regular intervals, e.g., once a week. Let $\bm{U} \in \mathbb{R}^{K\times n_b}$ represent the set of RSS values for this new batch, where $K$ is the number of new samples and $n_b$ is the total number of MACs detected in this batch. As done for the initial fingerprint database, if a MAC is not detected in a new sample, its corresponding RSS value is filled with $-120$ dBm. To ensure consistency, we also apply the same normalization process to each entry of $\bm{U}$.

\begin{figure}
    \centering
    \subfigure[Encoder]{\includegraphics[width=0.13\linewidth, angle=90]{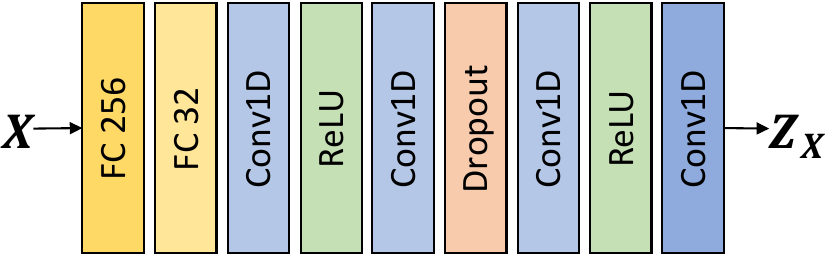}} 
    \subfigure[Decoder]{\includegraphics[width=0.13\linewidth, angle=90]{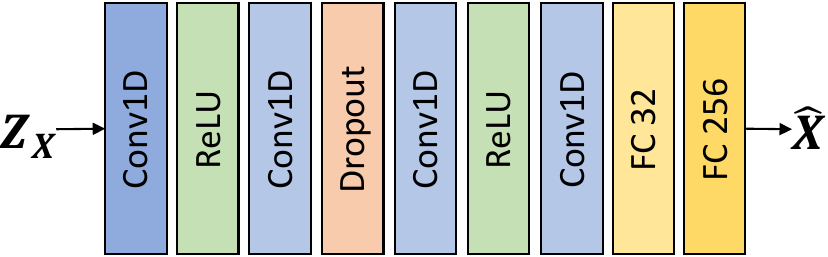}}
    \caption{Structure of the autoencoder.}
    \label{fig:ae}
    \Description[Autoencoder]{Showing the general structure of \n{}'s autoencoder, including one encoder and one decoder.}
\end{figure}

To reduce the dimensions of RSS values and extract fixed-length feature vectors for representing the samples in the initial fingerprints as well as from new batches of samples, we employ an autoencoder as our feature extractor. The autoencoder comprises an encoder $\mathbb{E}(\cdot)$ and a decoder $\mathbb{G}(\cdot)$, and its architecture is illustrated in Figure~\ref{fig:ae}. For the RSS-value set $\bm{X}$, we generate its corresponding feature set $\bm{Z}_X$ by passing it through the encoder $\mathbb{E}(\cdot)$. This operation can be expressed as
\begin{equation}
    \bm{Z}_X =\mathbb{E}(\bm{X}).
\label{eqn:z_x}
\end{equation}
The extracted feature set $\bm{Z}_X$ then goes through the decoder $\mathbb{D}(\cdot)$ to generate the following recovered RSS set: 
\begin{equation}
    \hat{\bm{X}} = \mathbb{D}(\bm{Z}_X).
\label{eqn:decoder}
\end{equation}
The autoencoder network is trained by minimizing the following loss of the feature extraction:
\begin{equation*}
    \mathcal{L}_F = \| \bm{X} - \hat{\bm{X}} \|_F,
\end{equation*}
where $\|\cdot\|_F$ is the Frobenius norm.

Once the autoencoder is trained, we obtain the output feature set $\bm{Z}_X$, which represents the features of the signals in the fingerprint database. When a new batch of WiFi signals $\bm{U}$ is collected, we can extract their features by employing the same encoder. This operation can be expressed as
\begin{equation}
\bm{Z}_U=\mathbb{E}(\bm{U}).    
\label{eqn:z_u}
\end{equation}

\subsection{Graph Formulation and GNN Training}
\label{subsec:graph_training}
Given the RSS-value set $\bm{X}$ and its corresponding extracted feature set $\bm{Z}_X$, we construct a graph $\mathcal{G}=(\bm{V}, \bm{E}, \bm{W})$, where $\bm{V}$ represents the nodes, $\bm{E}$ denotes the edges, and $\bm{W}$ contains the corresponding edge weights. For each sample $\bm{x}$ in $\bm{X}$, which is a row vector, we model it as a sample node $\bm{v}_x$ in the graph. Additionally, we model each detected MAC address in $\bm{X}$ as an AP node. If an AP node $\bm{v}_m$ is detected in the sample $\bm{x}$, its corresponding node in the graph is the sample node $\bm{v}_x$. Consequently, an edge $\bm{e}_{xm} \in \bm{E}$ is established between $\bm{v}_x$ and $\bm{v}_m$. Its edge weight is defined as a non-negative function of the detected RSS value. Specifically, the edge weight $\bm{w}_{xm}$ is calculated as $\bm{x}_m + c$, with $c > \max\{|\bm{X}_{ij}|, \forall i,j\}$, where $\bm{x}_m$ represents the RSS value of AP $m$ in $\bm{x}$, and $\bm{X}_{ij}$ denotes the RSS value of AP $j$ in WiFi signal sample $i$. We here set $c$ to 120, which is a common practice for adding RSS values to graphs in~\cite{zhuo2022grafics, zhang2020graph, zhang2023dagraph, zhuo2023fisoneflooridentificationlabel}.
Note that if a MAC address is not detected in a WiFi sample $\bm{x}$, its corresponding value is filled with $-120$ dBm. In such a case, no edge exists between the corresponding pair of nodes in the graph.

Each sample node $\bm{v}_x$ in the graph is initially associated with the feature vector of $\bm{x}$ obtained by the feature extractor in Section~\ref{subsec:RSS_extractor}. Letting $\bm{z}_x$ be the feature vector of $\bm{x}$, it is a row vector in $\bm{Z}_X$. We augment this $\bm{z}_x$ by concatenating $\bm{z}_x$ with its location label $\bm{y}$. For simplicity, we continue to denote the augmented vector as $\bm{z}_x$ throughout the rest of the paper. Then for any AP node $\bm{v}_m$, its node feature is initialized as the weighted average of the node features in its immediate neighborhood $\mathcal{N}(\bm{v}_m)$, which are all sample nodes. This initialization reflects the spatial relationships among the nodes and is given by
\begin{equation*}
    \bm{z}_m=\sum_{v_x\in\mathcal{N}(\bm{v}_m)}\frac{\bm{w}_{mx}}{\sum_{v_y\in\mathcal{N}(\bm{v}_m)} \bm{w}_{my}}\bm{z}_{x}.
\end{equation*} This approach effectively aggregates information from all neighboring sample nodes of each AP node, incorporating that information into the AP nodes.

\begin{figure*}
    \centering
    \subfigure[]{\includegraphics[width=0.38\textwidth]{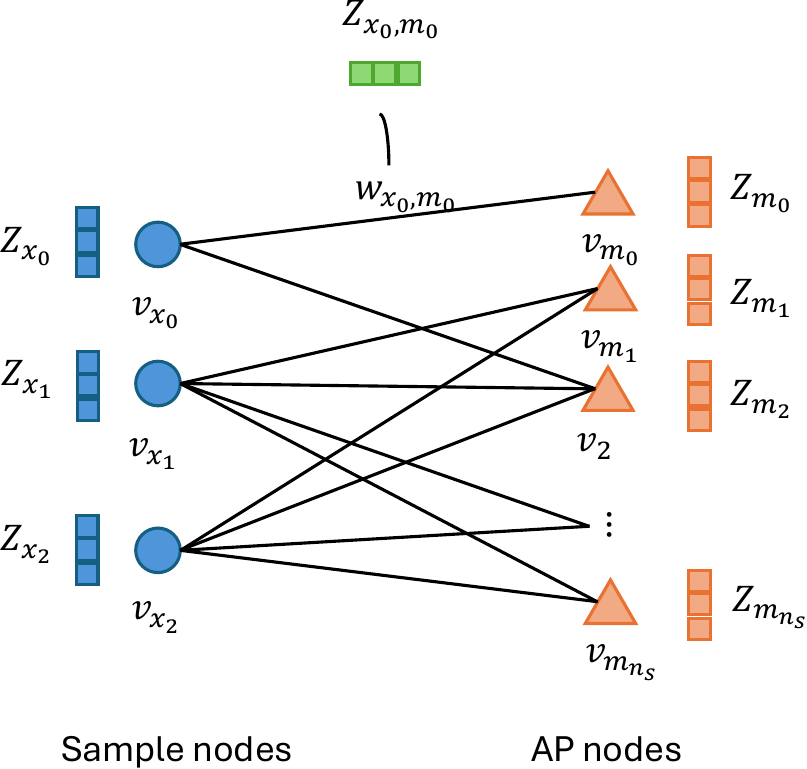}} 
    \hspace{0.02in}
    \subfigure[]{\includegraphics[width=0.48\textwidth]{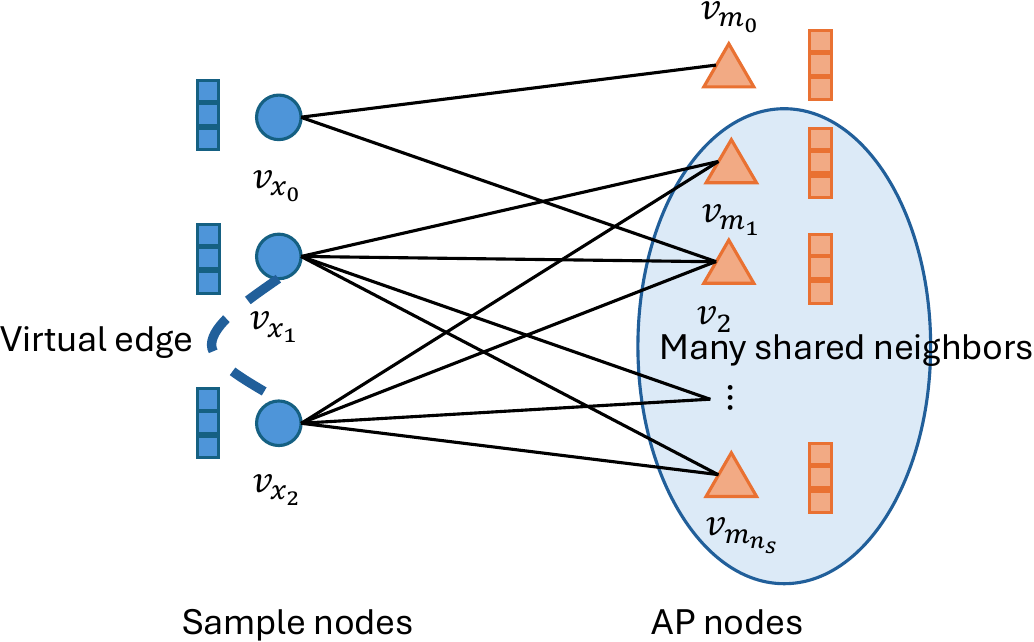}}
    \caption{Virtual edge creation. (a) Before virtual edge creation, edges only exist between sample nodes $\bm{v}_x$'s and AP nodes $\bm{v}_m$'s; (b) After creation, virtual edges are added among sample nodes having many shared AP neighbors like $\bm{v}_{x_1}$ and $\bm{v}_{x_2}$.}
    \label{fig:virtual_edge}
    \Description{Illustration of virtual edge creation: (a) shows edges between sample nodes and AP nodes only, while (b) shows additional virtual edges among sample nodes with shared AP neighbors.}
\end{figure*}

Thus far, edges have only been established between sample nodes and AP nodes. However, to update the existing fingerprints, which consist of signal samples, it is important to emphasize the sample nodes that contain features for these signals. To enable direct interactions between similar signal samples, we create virtual edges, denoted as $\bm{E}_{virtual}$, connecting these sample nodes. The similarity between signal samples is determined by the common AP neighbors that they share, as illustrated in Figure~\ref{fig:virtual_edge}. This similarity can be quantified using cosine similarity between each pair of sample nodes, which is calculated as the dot product of their respective node feature vectors.

\begin{figure}
    \centering
    \includegraphics[width=0.65\textwidth]{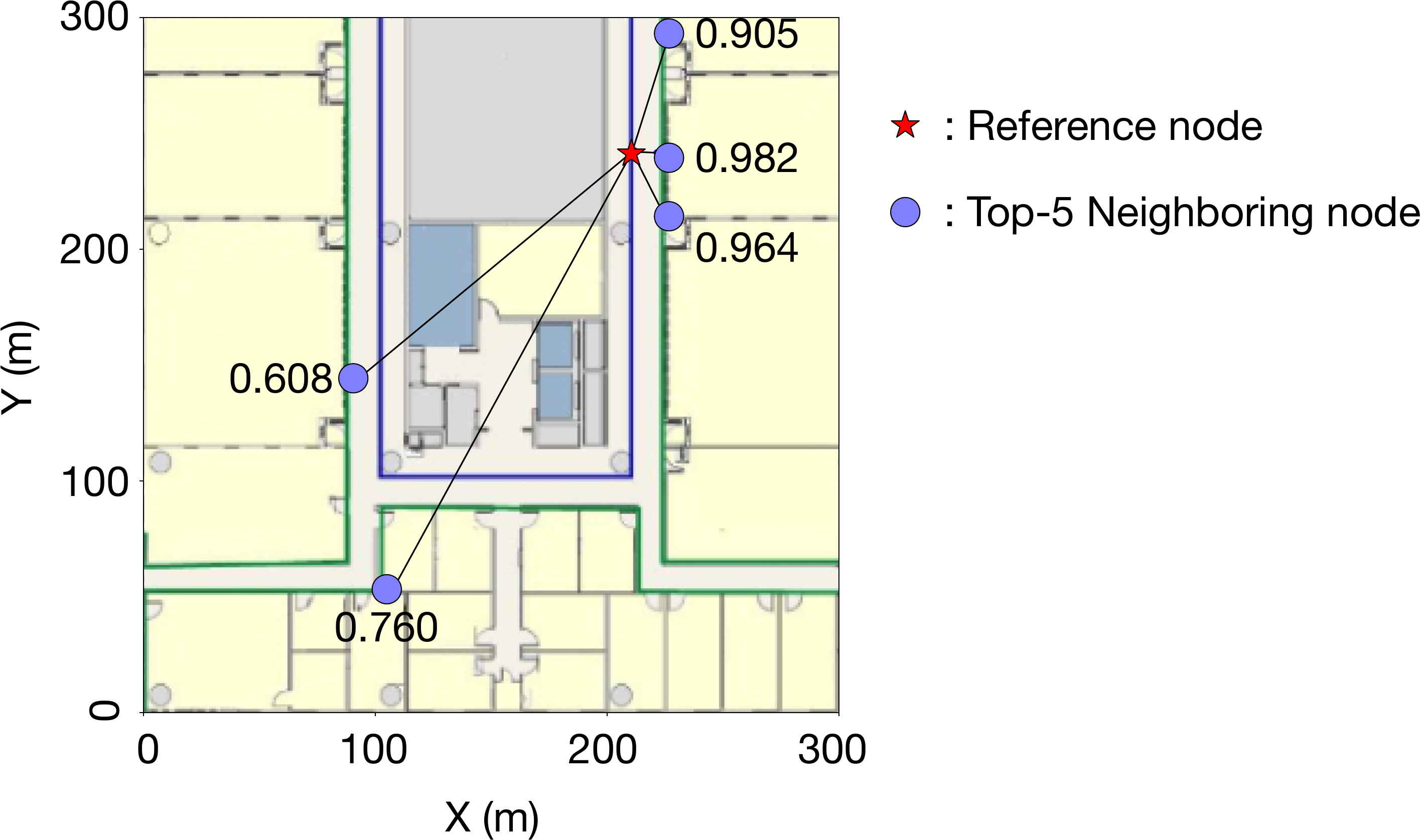}
    \caption{Feature similarity between a reference node and its neighboring nodes can be low even for top-5 neighbors.}
    \label{fig:top_k}
    \Description{Illustration of the issue of top-k neighbors.}
\end{figure}

To enable direct connections between similar sample nodes, one potential approach is to identify the $k$ nearest neighbors of each node based on their cosine similarity values. However, it remains a challenge how to determine the appropriate value of $k$, as a neighboring node with low similarity could still appear in the top-$k$ list regardless of the choice of $k$. 
This problem is illustrated in Figure~\ref{fig:top_k}, where the top five neighbors of a WiFi sample node are identified, but two of them display notably low similarity. To mitigate this problem, \n{} creates virtual edges using a thresholding method with a predefined threshold $\sigma$ on the cosine similarity values. In this process, only pairs of sample nodes with similarity values greater than $\sigma$ can be connected by virtual edges.

To learn a node embedding (or an updated node feature vector) for each node in the constructed graph $\mathcal{G}$, we construct a GNN by employing its training process based on GraphSAGE \cite{hamilton2017graphsage}. The principle behind GraphSAGE is that the embedding of each node is learned by aggregating information from its sampled one-hop neighborhood, e.g., calculating the mean value of the node embeddings within the neighborhood. By employing $L$ layers of aggregation, the embedding of each node incorporates information from its $L$-hop neighbors. In \n{}, we adapt and modify the aggregation process as follows. 

We first introduce an edge feature vector as an additional input to the aggregation process. For each edge $\bm{e}_{uv} \in \bm{E} \cup \bm{E}_{virtual}$, we initialize its edge feature vector $\bm{z}_{uv}$ as the mean value of the node features of its connected end nodes $u$ and $v$, i.e., $\bm{z}_{uv} = (\bm{z}_u + \bm{z}_v)/2$. The node-wise and edge-wise features are then fed into the aggregation process. The aggregation process is summarized in Algorithm~\ref{alg:init}, where $L$ denotes the number of aggregation layers, and $v\_{edge}(\cdot)$ represents the creation of virtual edges. Before the $L$ layers of aggregation, virtual edges are constructed. The process at each layer can be divided into the following three steps:

\begin{algorithm}
{
{\small
\caption{\n{}: Training}\label{alg:init}

\textbf{Input:} Graph $G(V, E, W)$; initialized node embeddings $\bm{z}_{v}$, $\forall v\in V$; number of aggregation layers $L$.

\textbf{Output:} Node embeddings $\bm{z}_v, \forall v\in V$.

\nl  $E_{virtual}=v\_edge(E, V, \sigma)$. \hfill \Comment{Create virtual edges between sample nodes.}

\nl $\bm{z}_{uv}^0\gets \frac{\bm{z}_u+\bm{z}_{v}}{2},\ \forall e_{uv}\in E\cup E_{virtual}$.

\nl \For{$l=1, 2, \cdots, L :$}{

\nl     \For{$v\in V$} {

\Comment{Aggregate node features using the features from the previous layer.}
\nl        $\bm{z}_{hv}^{l}=AGG\left(\phi(\mathbb{W}_0^{l}\cdot \text{CONCAT}(\bm{z}_{uv}^{l-1}, \bm{z}_u^{l-1})), ~\forall e_{uv }\in E\cup E_{virtual}\right)$, 

\Comment{Update node features.}        
\nl        $\bm{z}_v^{l}=\phi\left(\mathbb{W}_1^{l}\cdot \text{CONCAT}(\bm{z}_{v}^{l-1}, \bm{z}_{hv}^{l})\right)$

\nl        $\bm{z}_v^{l}=\frac{\bm{z}_v^{l}}{||\bm{z}_v^{l}||_2}$ 
        }
        
\nl     \For{$e_{uv}\in E$} {
\Comment{Update edge features.}
\nl            $\bm{z}_{uv}^{l}=\phi\left(\mathbb{W}_2^{l}\cdot \text{CONCAT}(\bm{z}_{u}^{l},\bm{z}_{uv}^{l-1},  \bm{z}_{v}^{l})\right)$

\nl         $\bm{z}_{uv}^{l}=\frac{\bm{z}_{uv}^{l}}{||\bm{z}_{uv}^{l}||_2}$ 
        }        
    }
\nl \Return{$\bm{z}_v, \forall v\in V$.}
}
}
\end{algorithm}

For every node $v$, we first aggregate the embeddings (or feature vectors) from its sampled neighbors, where the embedding of each neighbor is concatenated with the embedding of its connected edge (Line 5). Subsequently, the aggregated embedding from the neighbors is concatenated with the node's original embedding (Line 6). This concatenated embedding is then normalized (Line 7) to yield a new node embedding. At the end of each layer, the embedding of each edge is updated by concatenating the new embeddings of its end nodes with its existing edge embedding (Line 9 and Line 10). The learnable weights $\mathbb{W}_0^{l}$, $\mathbb{W}_1^{l}$, and $\mathbb{W}_2^{l}$ are trained by minimizing the difference between nodes and their immediate neighbors in terms of their embeddings. In other words, the training loss is given by
\begin{equation}
\mathcal{L}_{\mathcal{G}} = -\sum_{e_{uv}\in E\cup E_{virtual}}\log(\phi(\bm{z}_u^T\bm{z}_v)),
\label{eqn:loss_graph}
\end{equation}
where $\phi(\cdot)$ is an activation function. After the aggregation, to reduce the space required for maintaining the graph $\mathcal{G}$, the virtual edges are removed. This completes the initialization of \n{}.

\section{FINGERPRINT UPDATE FOR EXISTING APS}
\label{sec:update}
In this section, we present the batch update of \n{} during the online stage. This update process involves the existing APs that occur in the existing fingerprint database. These APs include the shared APs, which are common to both the new batch of signals and the existing fingerprint database, as well as those that are missing from the new batch. The trained GNN processes the new signals and generates their corresponding embeddings (Section~\ref{subsec:graph_inference}). Using those embeddings, an MLP-based update module is utilized for location predictions and fingerprint updates (Section~\ref{subsec:loc_predictor}).

\subsection{GNN Feature Aggregation}
\label{subsec:graph_inference}

When a new batch of signals $\bm{U}$ arrives, new sample nodes are created based only on the RSS values from the existing APs and added into $\mathcal{G}$. The initial node features of these new sample nodes, obtained by the trained encoder in Section~\ref{subsec:RSS_extractor}, are represented by $\bm{Z}_U$. Each column vector in $\bm{Z}_U$ corresponds to the embedding of a new sample node, which is then concatenated with a randomly assigned two-dimensional location label. This concatenation is the same as what was done in Section~\ref{subsec:graph_training}. 

Let $V_U$ represent the set of new sample nodes, and let $V_X$ denote the set of existing sample nodes. Following the process in Section~\ref{subsec:graph_training}, a new set of virtual edges is created between all sample nodes in $V_U\cup V_X$ using their current node features. Subsequently, \n{} retrieves the updated node features for all sample nodes and existing AP nodes by executing Algorithm~\ref{alg:init}. In other words, the embeddings of nodes in $V_U\cup V_X$ and all relevant edge features are updated via the aforementioned aggregation process. In particular, the embeddings of new sample nodes, i.e., $\bm{Z}_U$, are used to update the existing fingerprint database.

\subsection{Update Module}
\label{subsec:loc_predictor}

To iteratively predict the locations of new samples and obtain updated features for existing samples in the fingerprint database, we introduce an update module consisting of two sequential MLP networks. For a new batch of samples, the first MLP network takes their feature vectors $\bm{Z}_U$ as input and predicts their corresponding location labels $\hat{\bm{Y}}_U$. In addition, the second MLP network is responsible for taking the location labels of samples in the existing fingerprint database and producing their updated feature vectors $\hat{\bm{Z}}_X$. Here we utilize $\hat{\bm{Z}}_X$ to recover their corresponding \emph{updated} RSS values $\hat{\bm{X}}$. This process is achieved using the decoder as in Equation~\eqref{eqn:decoder}. The updated RSS values are then used to replace the old ones in the fingerprint database. After this update is completed, the new sample nodes, along with their connected edges, are removed from the graph. Consequently, the number of nodes of the graph $\mathcal{G}$ remains unchanged. In other words, each batch of new samples is utilized to update the fingerprint database without expanding its size.

The first MLP network is trained using the feature vectors of existing samples $\bm{Z}_X$ and their corresponding location labels $\bm{Y}_X$ in the fingerprint database. Specifically, the network is trained to minimize the following loss function, which measures the discrepancy between the predicted location labels $\hat{\bm{Y}}_X$ and the true labels $\bm{Y}_X$:
\begin{equation}
\mathcal{L}_\mathcal{P}=||\bm{Y}_X-\hat{\bm{Y}}_X||_F.
\end{equation}
Similarly, the second MLP network is designed to minimize the following loss function, which measures the discrepancy between the updated feature vectors $\hat{\bm{Z}}_U$ (the output of the MLP network) and their corresponding ground truth vectors $\bm{Z}_U$:
\begin{equation}
\mathcal{L}_\mathcal{U}=||\bm{Z}_U-\hat{\bm{Z}}_U||_F.
\end{equation}

It is worth noting that the two MLP networks are similar to an autoencoder. The first MLP network serves as the encoder, which is responsible for the location prediction. The second MLP network acts as the decoder, which is for the fingerprint update.

In addition to $\mathcal{L}_\mathcal{P}$ and $\mathcal{L}_\mathcal{U}$, we define a consistency loss among neighboring nodes in the graph to train the MLP networks with higher accuracy. The rationale behind this consistency loss is that the newly predicted location or updated feature vector of each node should not differ too much from its neighborhood. It has two parts. The first one is for the location prediction. The predicted locations of new sample nodes should have similar location labels to those of old sample nodes in their virtual neighborhoods. Thus, we define the following loss function:
\begin{equation}
\mathcal{L}_{CP}=||\bm{Y}_X\bm{A}-\bm{Y}_U||_F,
\end{equation}
where $\bm{A}$ is a matrix with elements $\bm{A}_{ij}$ being $\bm{A}_{ij} = 1$ for all $e_{uv}\in E_{virtual}$ and $\bm{A}_{ij} = 0$ otherwise. The second part is for the feature-vector update. Similar to the first part, the updated feature vectors of (old) sample nodes in the database should be similar to the features of new sample nodes in their virtual neighborhoods, leading to the following loss function:
\begin{equation}
\mathcal{L}_{CU}=||\bm{Z}_X\bm{A} - \bm{Z}_U||_F.
\end{equation}

To summarize, considering all four loss functions, we have the final loss function to train the update module for each batch of new samples, which is given by 
\begin{equation}    \mathcal{L}=\alpha(\mathcal{L}_{\mathcal{P}}+\mathcal{L}_{\mathcal{U}})+(1-\alpha)(\mathcal{L}_{CP}+\mathcal{L}_{CU}),
    \label{eqn:loss_mlp}
\end{equation}
where $\alpha$ is the regularization parameter.

After applying the MLPs for signal and location updates, we remove the virtual edges created for this batch of new signals to minimize the storage requirements for the graph $\mathcal{G}$, similar to the approach described in Section~\ref{subsec:graph_training}. Additionally, for every batch update, we update the trainable weights of the GNN by minimizing the loss function in Equation~\eqref{eqn:loss_graph}. This loss function captures the differences between the embeddings of recently created nodes and their neighbors, including both direct neighbors and those connected through virtual edges. Furthermore, since $\bm{X}$'s feature vectors $\bm{Z}_X$ are updated for each new batch and the corresponding APs may changed, the original autoencoder should also be updated to ensure that $\bm{Z}_X$ can be extracted from $\bm{X}$. To resolve this, we retrain it using the following training loss:
\begin{equation}
    \mathcal{L}_F = \| \bm{X} - \hat{\bm{X}} \|_F+\frac{1}{2}(\| \bm{\mathbb{E}(X)} - \bm{Z}_X\|_F+\|\mathbb{D}(\bm{Z}_X)-\hat{\bm{X}} \|_F).
\end{equation}

\section{FINGERPRINT UPDATE WITH AP CHANGES}
\label{sec:update2}
In this section, we explain how \n{} updates the edges in the graph $\mathcal{G}$ according to AP changes, including the addition of new APs and the removal of existing ones. For a new batch of signals, \n{} first assesses whether any new AP nodes need to be created. If so, it creates the new AP nodes and utilizes the node features from the graph whose AP nodes only consist of the existing APs to determine where new edges should be established between the new AP nodes and existing sample nodes (Section~\ref{subsec:new_ap_pred}). \n{} also evaluates whether any existing edges should be removed from the graph, regardless of the presence of new APs (Section~\ref{subsec:ap_removal}). Throughout this process, the node features remain unchanged.

\subsection{Edge Prediction for New APs}
\label{subsec:new_ap_pred}

After updating the node features for the existing and new sample nodes $V_U\cup V_X$ using only the existing APs, \n{} checks for any APs, identified by their MAC addresses, that are detected in $\bm{U}$ but not present in $\bm{X}$. If such APs are identified, new corresponding AP nodes are created in the graph $\mathcal{G}$ for those MAC addresses.  Then, similar to the initial training process in Section~\ref{subsec:graph_training}, the initial feature vector for each newly added AP node is calculated as the weighted average of the node feature vectors of its immediate neighbors.

To effectively utilize incoming new samples for updating the signals in the existing fingerprint database—especially the RSS values from new APs—we introduce an edge prediction algorithm. This algorithm identifies potential missing connections between AP nodes and sample nodes, making it particularly useful for newly detected AP nodes and existing sample nodes.

\begin{figure}
    \centering
    \includegraphics[width=0.5\textwidth]{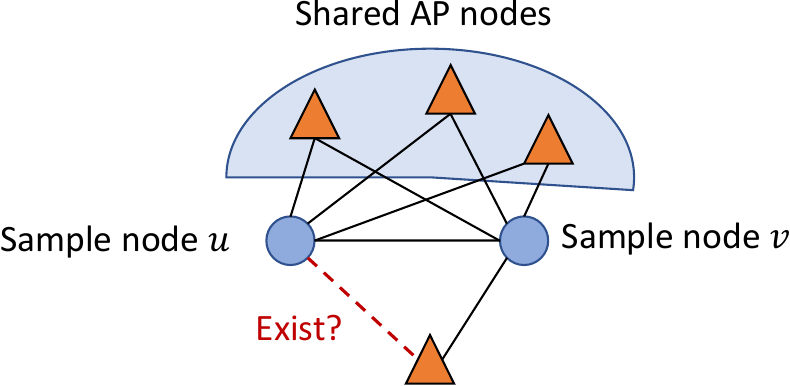}
    \caption{Edge prediction.}
    \label{fig:link_prediction}
    \Description{Illustration of edge prediction between new AP nodes and existing sample nodes.}
\end{figure}

As depicted in Figure~\ref{fig:link_prediction}, our edge prediction algorithm aims to predict missing edges between pairs of AP nodes and sample nodes that are not currently connected but are both present in the neighborhood of another sample node. As mentioned in Section~\ref{subsec:graph_training}, virtual edges are established between similar sample nodes. Hence, we search for node pairs consisting of an AP node and a sample node within the neighborhood of sample nodes that serve as endpoints of virtual edges. This edge prediction process allows us to discover potentially close AP nodes and sample nodes that are not yet connected in the graph.

In the algorithm, we adapt two metrics proposed in~\cite{kumar2016edge}, namely, goodness and fairness, to assess edge weights in the graph and decide whether an AP node and a sample node should be connected. The goodness metric indicates how much a node is trusted by its neighbors as a similar node, while the fairness metric measures how reliable a node is in evaluating the goodness of its neighbors. It is important to note that goodness and fairness are interdependent. Specifically, if a node exhibits high goodness and high fairness, nearby nodes are more likely to connect with it.

\begin{figure}
    \centering
    \includegraphics[width=0.6\linewidth]{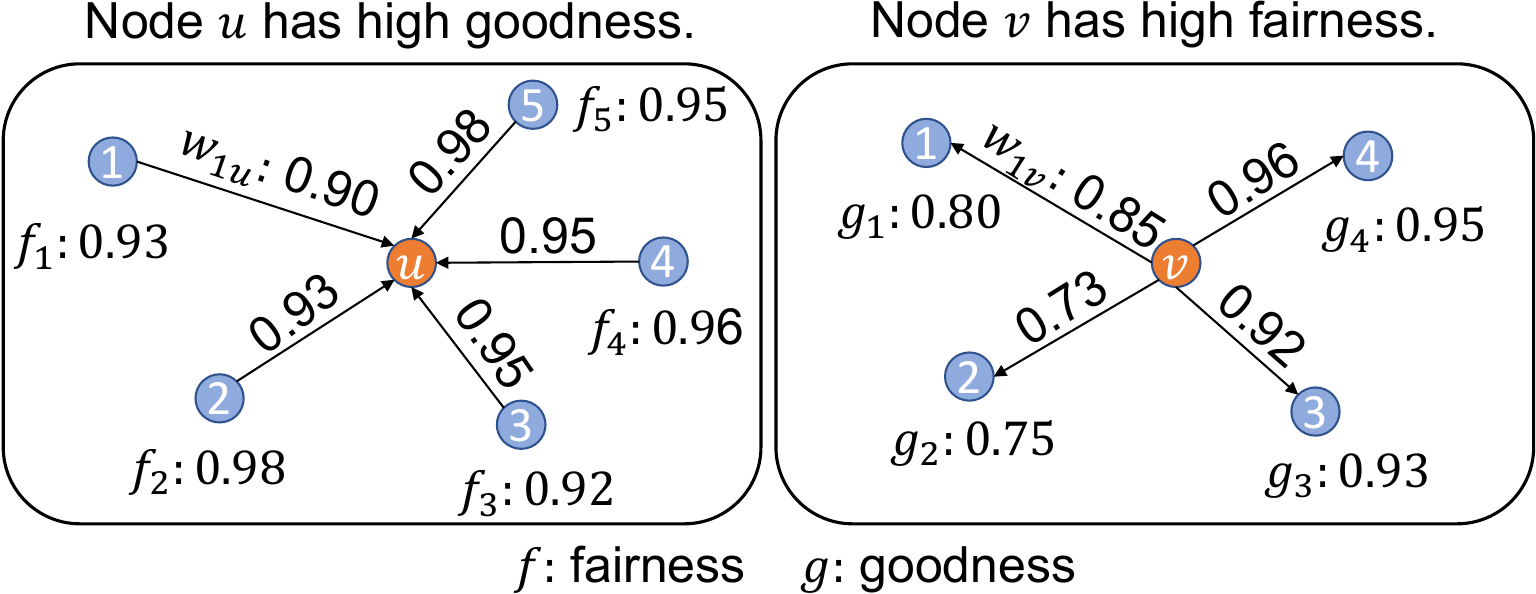}
    \caption{An illustrative example of goodness $g$ and fairness $f$. }
    \label{fig:goodness_fairness}
    \Description{An illustrative example of goodness $g$ and fairness $f$.}
\end{figure}

To define the two metrics explicitly in our scenario, we define the goodness $\bm{G}(v)$ of node $v$ as the normalized weighted average of its neighbors' fairness values. The higher the value, the better the goodness. We also define the fairness $\bm{F}(v)$ of node $v$ as the normalized difference between the weights of its connected edges and the goodness values of its neighbors. The larger the difference is, the lower its fairness. We provide an illustrative example of goodness and fairness in Figure~\ref{fig:goodness_fairness}, where in our case the goodness and fairness are quite close to the edge weights. As illustrated in the figure, node $u$ is considered good as the weighted average fairness of its neighboring nodes is close to 1. Node $v$ is considered fair as the average difference in edge weights between it and its neighboring nodes is close to 0. As each node is associated with its $d$-dimensional embedding (feature vector), we define the goodness and fairness along each dimension, i.e., $\bm{G}(v) = [\bm{g}_1(v), \cdots, \bm{g}_d(v)]$, and $\bm{F}(v) = [\bm{f}_1(v), \cdots, \bm{f}_d(v)]$. Specifically, for each dimension $i$, we define $\bm{g}_i(v)$ and $\bm{f}_i(v)$ as
\begin{eqnarray}
    \bm{g}_i(v)&=&\frac{1}{|N(v)|}\sum_{u\in N(v)}\bm{f}_i(u)\overline{w}_{uv}, 
    \label{eqn:goodness}\\
    \bm{f}_i(v)&=&1-\frac{1}{2|N(v)|}\sum_{u\in N(v)}|\overline{w}_{uv} -\bm{g}_i(u)|,
    \label{eqn:fairness}
\end{eqnarray}
respectively, where $\overline{w}_{uv}$ is the normalized weight of edge $e_{uv}$, which is defined as $\overline{w}_{uv} = w_{uv}/w_{\max}$, with $w_{\max} = \max\{w_{ij}, \forall i,j\}$.

\setlength{\textfloatsep}{10pt}
\begin{algorithm}[t]
{
{\small
\caption{\n{}: Edge Modification}
\label{alg:link_prediction}

\textbf{Input:} Graph $\mathcal{G}(V, E \cup E_{virtual}, W)$; extracted node feature $\textbf{z}_{v}~,\ \forall v\in V$.

\textbf{Output:} New edges $E_n$ and corresponding edge weights $W_n$.

\nl $\overline{w}_{uv}=\frac{w_{uv}}{w_{\max}}, \forall e_{uv} \in E \cup E_{virtual}$.

\nl $\bm{G}^{(0)}(v)=\bm{F}^{(0)}(v)=\bm{z}_{v}$, $\forall v \in V$.

\nl $E_n = W_n = \{\}$

\nl $t=0$.

\nl \Do{$\sum_{v\in V}\|\bm{F}^{(t)}(v)-\textbf{F}^{(t-1)}(v)\|_2>\epsilon$ and $\sum_{v\in V}\|\bm{G}^{(t)}(v)-\textbf{G}^{(t-1)}(v)\|_2>\epsilon$}{
    \Comment{Iteratively update $\bm{G}$ and $\bm{F}$.}
\nl \For{$v\in V$} {
        \For{$i \in [1,d]$}{
\nl     $\bm{g}_i^{(t+1)}(v)=\frac{1}{|N(v)|}\sum_{u\in N(v)}\bm{f}_i^{(t)}(u)\overline{w}_{uv}$

\nl     $\bm{f}_i^{(t+1)}(v)=1-\frac{1}{2|N(v)|}\sum_{u\in N(v)}|\overline{w}_{uv}-\bm{g}_i^{(t+1)}(u)|$
        }
}
\nl     $t=t+1$.
    }
\nl \For {$s,u,v\in V:e_{su}\in E, e_{uv}\in E_{virtual}$} {
    \Comment{$s$ is connected to $u$ but not necessarily $v$. $u$ and $v$ are virtual neighbors.}
\nl     \If {$v \notin N(s)$} {
\nl     $\hat{w}_{sv}=\frac{1}{2}w_{\max}\left(\bm{G}(s)\cdot \bm{F}(v) + \bm{F}(s)\cdot \bm{G}(v)\right)$

\nl     \If{$\hat{w}_{sv} \geq \delta$ and $e_{sv}\not \in E$}{
        \Comment{Add an edge between $s$ and $v$.}
        \nl $E_n = E_n \cup \{e_{sv}\}$  \\      
        \nl $W_n = W_n \cup \{\hat{w}_{sv}\}$
}

\nl     \If{$\hat{w}_{sv} < \delta$ and $e_{sv} \in E$}{
        \Comment{Remove the edge between $s$ and $v$.}
        \nl $E_n = E_n \smallsetminus \{e_{sv}\}$  \\      
        \nl $W_n = W_n \smallsetminus \{w_{sv}\}$
}

}
}
\nl \Return $E_n$, $W_n$.
}
}
\end{algorithm}

Our edge prediction algorithm is summarized in Algorithm~\ref{alg:link_prediction}. We first normalize the edge weights by the maximum edge weight in the graph (Line 1). The goodness and fairness vectors are both initialized as the updated node features (Line 2). We then iteratively update the goodness and fairness values for each node (Lines 5--9) until they converge. We next compute the predicted edge weights for potential edges, i.e., edges between AP nodes and sample nodes that are not connected but are both present in the neighborhood of another sample node (Line 12). If the predicted edge weight is greater than a (predetermined) threshold, then a new edge is added to the graph with the corresponding edge weight (Lines 13--15). The threshold is given by $\delta = 120-\max\{|\bm{X}_{ij}|, \forall i,j\}$. By setting this threshold, we can ensure that edges would only be created when the predicted weights are greater than $120-c$, which equals $0$ as we follow the common practice in ~\cite{zhuo2022grafics, zhang2020graph, zhang2023dagraph} and set $c=120$. Specifically, edges will only be created for those predicted to have non-negative edge weights, guaranteeing that all created edges possess valid, positive weights.

\subsection{Forgetting Mechanism for Existing APs}
\label{subsec:ap_removal}

In the process of updating fingerprints, it is important to consider not only the installation of new APs but also the removal of existing ones from a site. When an AP is removed, its signals will no longer be recorded, necessitating their removal from the fingerprint database. To tackle this issue, we introduce an edge removal process in our edge prediction algorithm, as outlined in Algorithm~\ref{alg:link_prediction}. During this process, \n{} identifies predicted edge weights that fall below the threshold for edge existence, which is the same threshold used for adding edges. If an edge's predicted weight is below this threshold, it is removed from the graph (Lines 16–18). Furthermore, if any AP node becomes disconnected from the rest of the graph as a result of the edge removal process, it indicates that the corresponding AP has been removed. In such cases, the AP node is deleted from the graph, effectively removing its presence from the fingerprint database.

After the APs changes are recorded by edge changes in the graph, using the new edges and the new node features, we retrain the feature extractor according to the same procedure as mentioned before (Section \ref{subsec:RSS_extractor}). The feature vectors are again fed to the GNN for inference (Section~\ref{subsec:graph_inference}). The output embeddings are then used by the decoder in the feature extractor to obtain the updated RSS values (Section~\ref{subsec:loc_predictor}).

\setlength{\abovecaptionskip}{3pt plus 2pt minus 1pt}
\section{EXPERIMENTAL EVALUATION}
\label{sec:eva}
In this section, we experimentally evaluate \n{}. We first present our experiment settings in Section~\ref{subsec:exp} and then compare the performance of \n{} with state-of-the-art algorithms in Section~\ref{subsec:eva}. We further demonstrate the effectiveness of each system component of \n{} and the impact of the system parameters in Section ~\ref{subsec:ablation}.

\subsection{Experimental Setup}
\label{subsec:exp}

\begin{figure*}
    \centering
    \includegraphics[width=\textwidth]{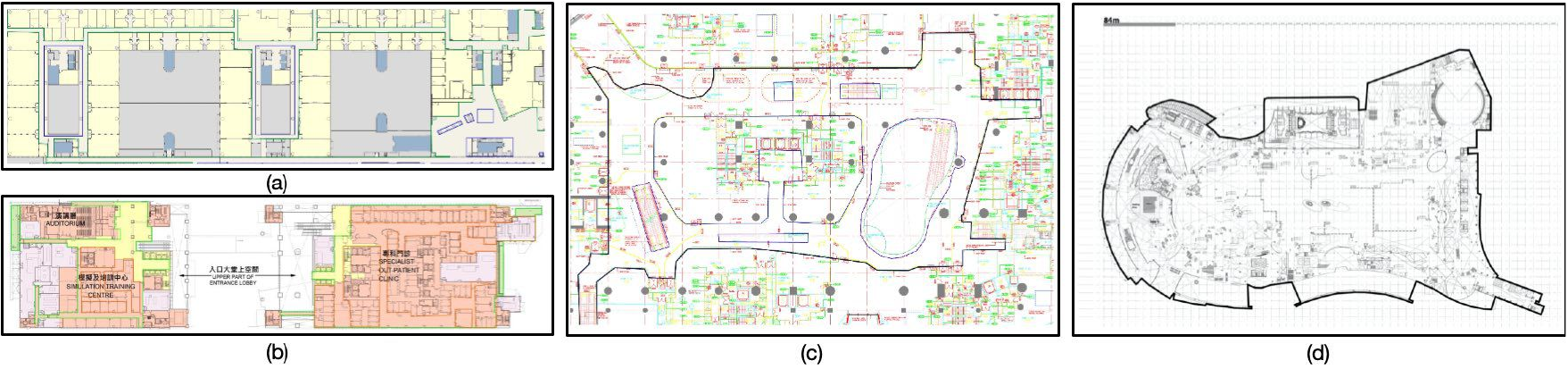}
    \caption{Floor plans of one floor from different sites. (a) Campus. (b) Hospital. (c) Mall A (d) Mall B.}
    \label{fig:maps}
    \Description{Four site maps.}
\end{figure*}

\noindent\textbf{Data collection:} We conduct experiments on four different sites, including one campus building (Campus), two shopping malls (Malls A and B), and one hospital (Hospital), as shown in Figure~\ref{fig:maps}. The campus building has three floors, and each of the other sites has four floors. For each floor in the buildings, we construct an initial fingerprint database by dividing the site into grids and collecting WiFi signals from the center of each grid. After that, WiFi signals are crowdsourced periodically for updates. For Campus and Mall A, we collect data once per week and once every two weeks, respectively, for eight weeks. For Mall B and Hospital, the data are collected every week for four weeks. For reliable performance evaluation, the measurement locations of the collected signals are also recorded. As a result, those locations can be used as ground truth, yet only for measuring the prediction error. 
In Table~\ref{tab:data}, we provide the details of the data collection process for fingerprint updates in four buildings. Additionally, to provide a clearer overview of the number of signal samples collected and the changes in the number of APs within each batch of new data, we present relevant statistics in Table~\ref{tab:summary_sample} and Table~\ref{tab:summary_aps}.

\begin{table}
\caption{Data collection details in different sites} 
\label{tab:data}
\begin{center}
\begin{tabular}{| c | c | c | c | c |} 
 \hline
Site & Area & Grid size & Duration & Frequency\\
  & (m$^2$) & (m$^2$)    &  (weeks)       & (/week) \\ 
 \hline
 Campus & $255$ $\times~ 95$ & $2.5$ $\times~ 2.5$ & 8 & 1 \\ 
 \hline
 Mall A & $180$ $\times~ 105$ & $2.5$ $\times~ 2.5$ & 8 & 0.5 \\
 \hline
 Mall B & $650$ $\times~ 280$ & $2.0$ $\times~ 2.0$ & 4 & 1 \\
 \hline
 Hospital & $200$ $\times~ 120$ & $2.0$ $\times~ 2.0$ & 4 & 1 \\
 \hline
\end{tabular}
\end{center}
\end{table}

\begin{table}
\caption{Number of samples collected on each floor for all sites}
\label{tab:summary_sample}
\begin{center}
\begin{tabular}{|c|c|c|c|c|c|c|c|c|c|c|}
 \hline
Site & Floor & Init & Week 1  & Week 2 & Week 3 & Week 4 & Week 5 & Week 6 & Week 7 & Week 8 \\ \cline{1-11}
\multirow{3}{*}{Campus} & GF & 11284 & 4029 & 4236 & 4661 & 4264 & 4342 & 4515 & 4509 & 4190 \\ \cline{2-11}
     & 1F & 10037 & 3589 & 3498 & 3981 & 3821 & 3445 & 3781 & 3448 & 3968 \\ \cline{2-11}
     & 2F & 11256 & 3733 & 3680 & 3667 & 3873 & 4110 & 3947 & 3899 & 3665 \\ \cline{1-11}
\multirow{4}{*}{Mall A} & B1F & 6284 & - & 2544 & - & 1788 & - & 2056 & - & 2272 \\ \cline{2-11}
     & GF & 7619 & - & 2487 & - & 2338 & - & 2843 & - & 2755 \\ \cline{2-11}
     & 1F & 3326 & - & 449 & - & 621 & - & 1054 & - & 965 \\ \cline{2-11}
     & 2F & 4158 & - & 1642 & - & 1153 & - & 1609 & - & 1629 \\ \cline{1-11}
\multirow{4}{*}{Mall B} & 1F & 131586 & 45417 & 46146 & 48439 & 46392 & - & - & - & - \\ \cline{2-11}
     & MF & 49453 & 16820 & 16394 & 16635 & 15660 & - & - & - & - \\ \cline{2-11}
     & 2F & 62882 & 21286 & 20567 & 20396 & 19944 & - & - & - & - \\ \cline{2-11}
     & 3F & 57194 & 52427 & 50251 & 53389 & 54803 & - & - & - & - \\ \cline{1-11}
\multirow{4}{*}{Hospital} & GF & 6154 & 2078 & 1733 & 1745 & 3936 & - & - & - & - \\ \cline{2-11}
     & 1F & 4318 & 258 & 419 & 419 & 1448 & - & - & - & - \\ \cline{2-11}
     & 2F & 2202 & 422 & 731 & 732 & 2305 & - & - & - & - \\ \cline{2-11}
     & 4F & 1931 & 194 & 194 & 118 & 263 & - & - & - & - \\ \cline{1-11}
\end{tabular}
\end{center}
\end{table}

\begin{table}
\caption{Summary of AP changes on each floor for all sites. +/- stands for addition/removal.}
\label{tab:summary_aps}
\begin{center}
\begin{tabular}{|c|c|c|c|c|c|c|c|c|c|c|c|}
 \hline
Site & Floor & Init & Week1  & Week2 & Week3 & Week4 & Week5 & Week6 & Week7 & Week8  & Total \\ 
  & & & +/- & +/- & +/- & +/- & +/- & +/- & +/- & +/-  & +/- \\ \cline{1-12}
\multirow{3}{*}{Campus} & GF & 391 & 53/46 & 33/26 & 29/22 & 3/2 & 41/6 & 17/28 & 25/23 & 3/4 & 204/136 \\ \cline{2-12}
     & 1F & 306 & 55/32 & 21/4 & 15/13 & 6/2 & 41/8 & 20/16 & 20/19 & 3/6 & 180/100 \\ \cline{2-12}
     & 2F & 253 & 10/7 & 19/5 & 12/9 & 0/4 & 27/3 & 5/8 & 11/7 & 6/13 & 90/56 \\ \cline{1-12}
\multirow{4}{*}{Mall A} & B1F & 927 & - & 17/10 & - & 43/27 & - & 58/43 & - & 55/21 & 140/68 \\ \cline{2-12}
     & GF & 1328 & - & 57/48 & - & 31/77 & - & 29/53 & - & 41/61 & 104/184 \\ \cline{2-12}
     & 1F & 434 & - & 18/24 & - & 37/38 & - & 24/37 & - & 11/34 & 47/90 \\ \cline{2-12}
     & 2F & 385 & - & 14/21 & - & 30/33 & - & 35/31 & - & 24/28 & 140/68\\ \cline{1-12}
\multirow{4}{*}{Mall B} & 1F & 494 & 63/41 & 18/52 & 15/27 & 6/4 & - & - & - & - & 82/104 \\ \cline{2-12}
     & MF & 199 & 3/1 & 5/4 & 2/2 & 0/3 & - & - & - & - & 8/8 \\ \cline{2-12}
     & 2F & 323 & 54/20 & 29/39 & 37/22 & 16/20 & - & - & - & - & 121/90 \\ \cline{2-12}
     & 3F & 304 & 18/30 & 33/62 & 50/39 & 14/13 & - & - & - & - & 64/93\\ \cline{1-12}
\multirow{4}{*}{Hospital} & GF & 673 & 8/4 & 5/8 & 5/0 & 4/1 & - & - & - & - & 17/8 \\ \cline{2-12}
     & 1F & 317 & 0/0 & 3/10 & 7/6 & 10/1 & - & - & - & - & 18/15 \\ \cline{2-12}
     & 2F & 266 & 0/0 & 11/7 & 2/0 & 3/0 & - & - & - & - & 15/6 \\ \cline{2-12}
     & 4F & 164 & 0/0 & 1/1 & 3/2 & 0/0 & - & - & - & - & 4/3\\ \cline{1-12}
\end{tabular}
\end{center}
\end{table}

\vspace{1mm}
\noindent\textbf{Implementation details:} In \n{}, for the autoencoder in feature extraction, we use a dropout rate of 0.5. The encoder has its input dimension as the number of MACs detected in the initial fingerprint database and its output dimension of 32. For the GNN in \n{}, we apply a learning rate of $0.01$ and a dropout of $0.5$. The model is trained for $50$ epochs. We use ReLU as the activation function $\phi$. We set $\sigma=0.95$ for the threshold value for creating virtual edges, $\alpha=0.5$ in Equation~\eqref{eqn:loss_mlp}, and $\epsilon=0.1$ for the link prediction threshold. The code is available anonymously online.\footnote{Code implementation: \url{https://github.com/khchiuac/GUFU}}

\vspace{1mm}
\noindent\textbf{State-of-the-art algorithms:} To evaluate the performance of \n{} on the fingerprint updates, we compare its performance with four state-of-the-art algorithms, namely, Fidora~\cite{chen2022fidora}, iToLoc~\cite{li2021imgadversarial}, MTDAN~\cite{wang2023mtdan} and WiDAGCN~\cite{zhang2023dagraph}. Each algorithm is briefly summarized as follows:
\begin{itemize}[itemsep=5pt,leftmargin=1.5em]

\item{Fidora}~\cite{chen2022fidora}: It uses a classification neural network and a reconstruction neural network to infer location labels for new signals and update RSS values in the original fingerprints via semi-supervised learning. It is designed based on the assumption that the signal characteristics, e.g., RSS values and detected APs, are different over different areas, which are grids in our experiments.

\item{iToLoc}~\cite{li2021imgadversarial}: It is based on the assumption that there are temporal features of WiFi signals that are consistent over time. Hence, it first models WiFi signals as a 2D image and extracts time-invariant features from the signals at different times using a convolutional time discriminator. It then predicts the locations of new signals using another convolutional neural network. Both networks are trained from the fingerprint database and the first batch of crowdsourced signals.   

\item{MTDAN}~\cite{wang2023mtdan}: Similar to our \n{}, MTDAN's update is also done with the assumption that there are AP differences in signal samples from different times. Utilizing multi-target domain adaptation to extract time-invariant signal features from stable APs (shared APs in our context), MTDAN is able to predict location labels for new signal samples. 

\item{WiDAGCN}~\cite{zhang2023dagraph}: It is one of the state-of-the-art graph modeling for signal fingerprint update. By modeling APs and signal samples as different nodes in the graph and using graph attention, WiDAGCN can match graphs constructed from new signals to subgraphs from the existing graphs, and thus get its signal and location predictions.   

\end{itemize}

Note that iToLoc~\cite{li2021imgadversarial} does not update the RSS values in the fingerprint database. For a fair comparison, we use our thresholding method (for creating virtual edges) in \n{} to do so, along with iToLoc. Specifically, for each signal record/sample in the database, which appears as a sample node in the graph, we first find its closest neighbors, which are also sample nodes in the graph, with their feature cosine similarity higher than $\sigma = 0.95$. Then, whenever the location of a new signal sample is predicted by iToLoc, its corresponding signal record/sample is found in the database, and each RSS value in the record is updated as the weighted average of the RSS values from its neighboring records/samples (i.e., its neighbors in the graph).

\noindent\textbf{Evaluation metrics:} To measure the accuracy of location prediction, we use the `location error' that is defined as the average Euclidean distance between the predicted locations and their ground-truth locations, as widely used in the literature. In addition, to measure how accurately the fingerprint database has been updated, we use the `RSS error' that is defined as the average difference between updated RSS values and actually measured RSS values of APs. All the experiments are done on a machine with an Intel(R) Core(TM) i9-9900X CPU @ 3.50GHz, 64G RAM and two graphic cards of Nvidia GeForce RTX 2080 Ti.

\begin{figure*}
    \centering
     \includegraphics[width=0.6\linewidth]{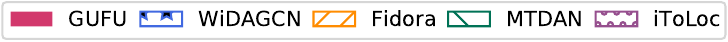}
            
        \subfigure[Campus]{
        \includegraphics[width=0.22\textwidth]{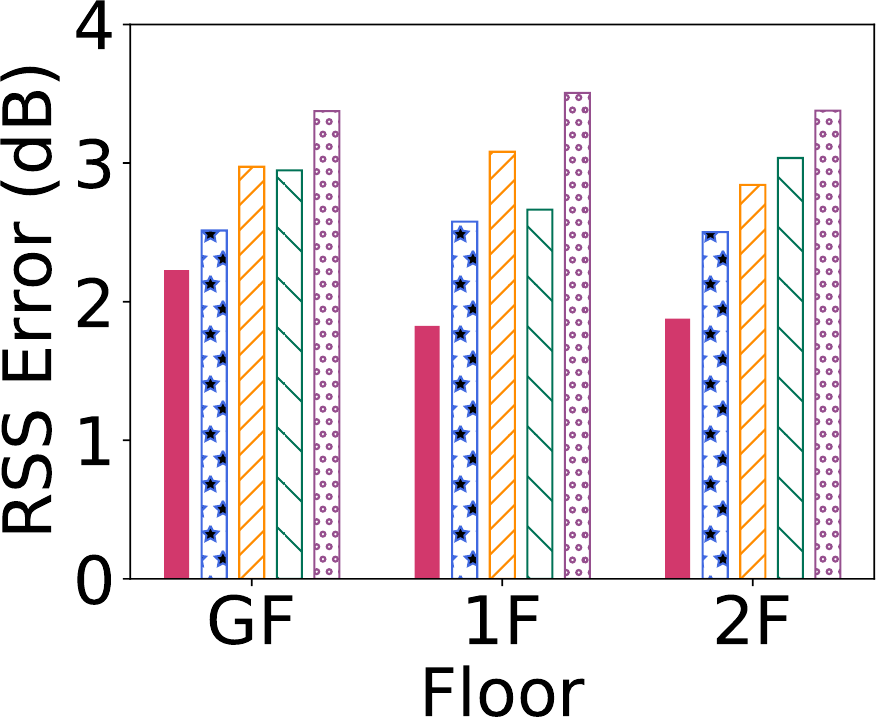}
        }            
        \subfigure[Mall A]{
        \includegraphics[width=0.22\textwidth]{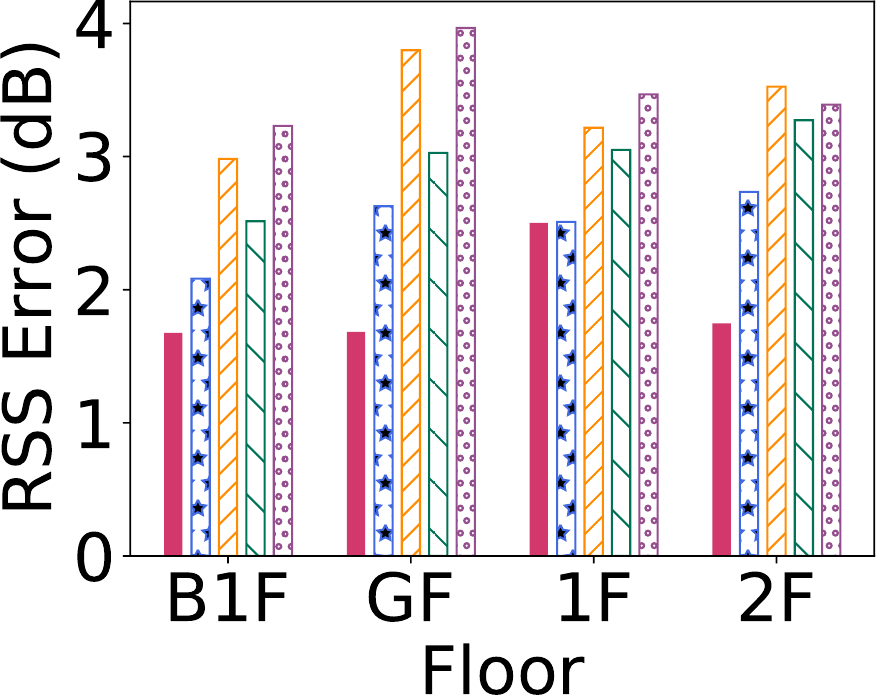}
        }
        \subfigure[Mall B]{
        \includegraphics[width=0.22\textwidth]{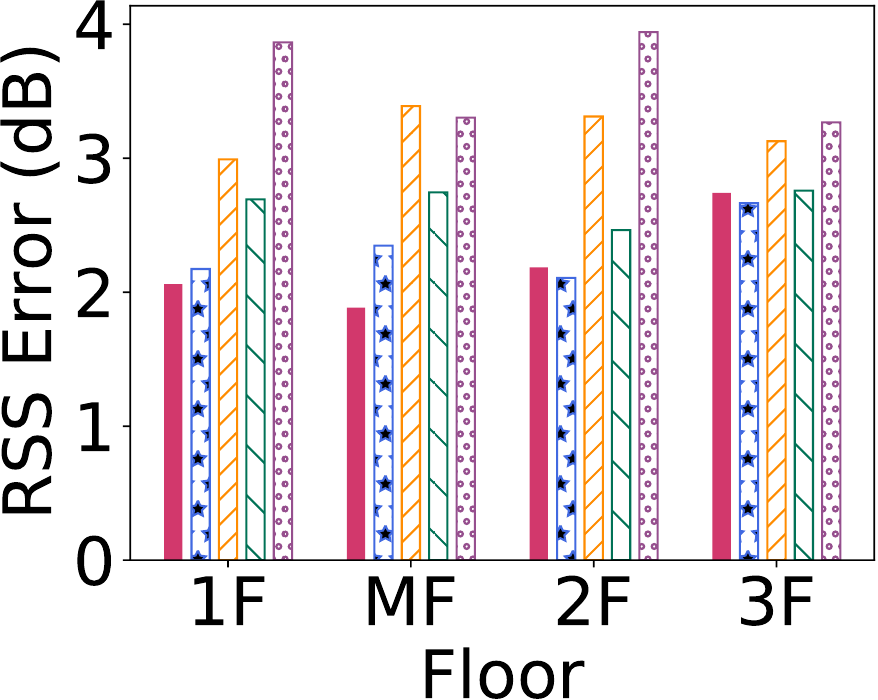}
        }
        \subfigure[Hospital]{
        \includegraphics[width=0.22\textwidth]{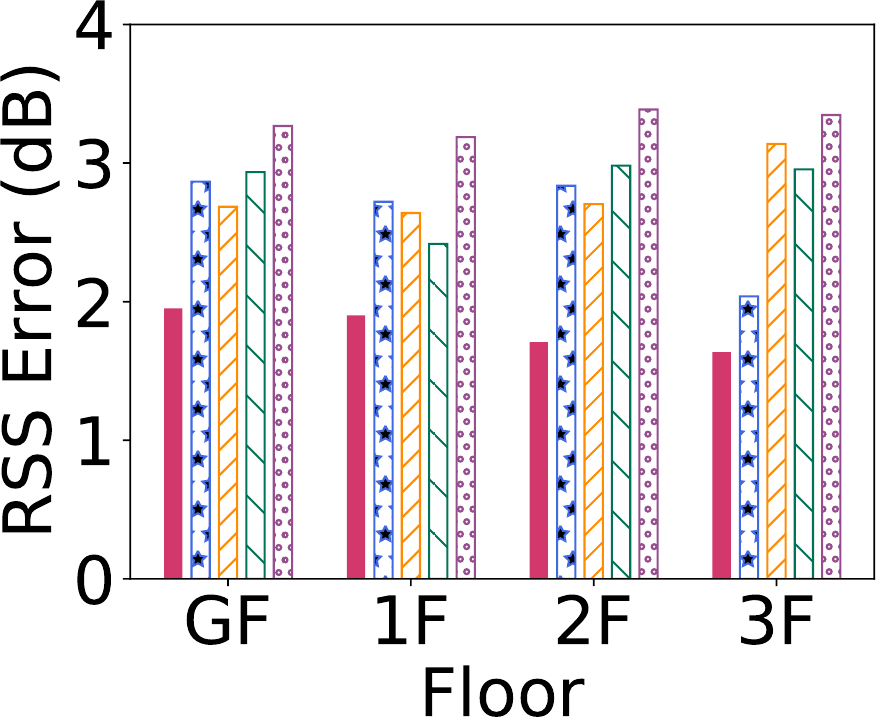}
        }    
    \caption{Summary of RSS error on different floors of the four sites.}
    \label{fig:exp2}
    \Description{RSS error, from exp1.}
\end{figure*} 

\subsection{Overall Performance}
\label{subsec:eva}

\begin{figure*}
    \centering
    \subfigure[GF]{\includegraphics[width=0.28\textwidth]{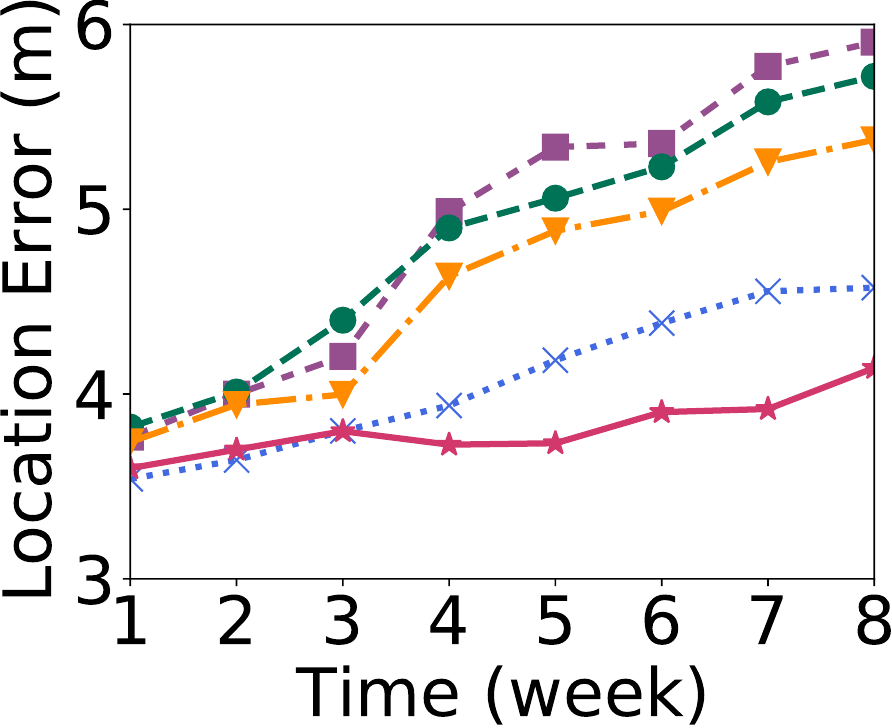}} 
    \hspace{0.02in}
    \subfigure[1F]{\includegraphics[width=0.28\textwidth]{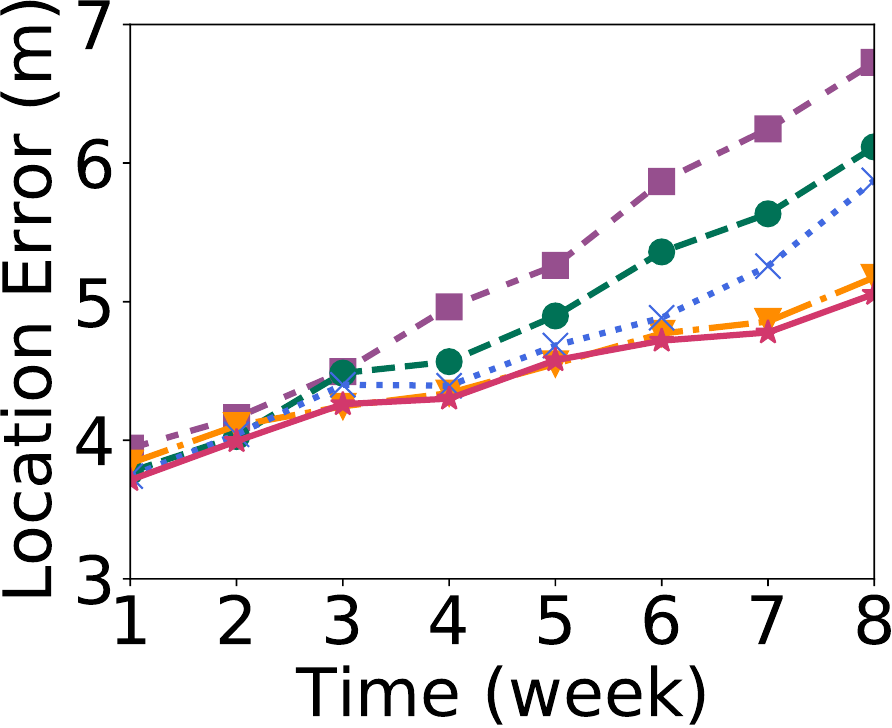}}
    \hspace{0.02in}
    \subfigure[2F]{\includegraphics[width=0.28\textwidth]{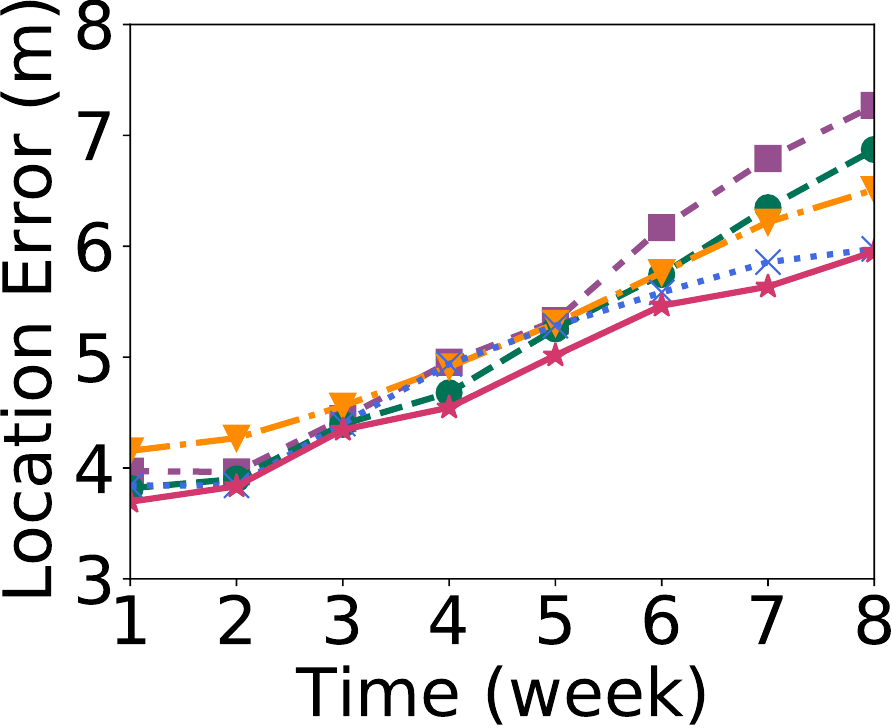}}
    \hspace{0.02in}
    \subfigure
    {\includegraphics[width=0.12\textwidth]{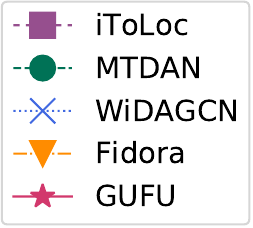}}
    \caption{Location error over eight weeks on three floors in the campus.}
    \label{fig:exp1_campus1}
    \Description{Loc error of campus1, from exp2.}
\end{figure*}

\begin{figure*}
    \centering
    \subfigure[B2F]{\includegraphics[width=0.21\textwidth]{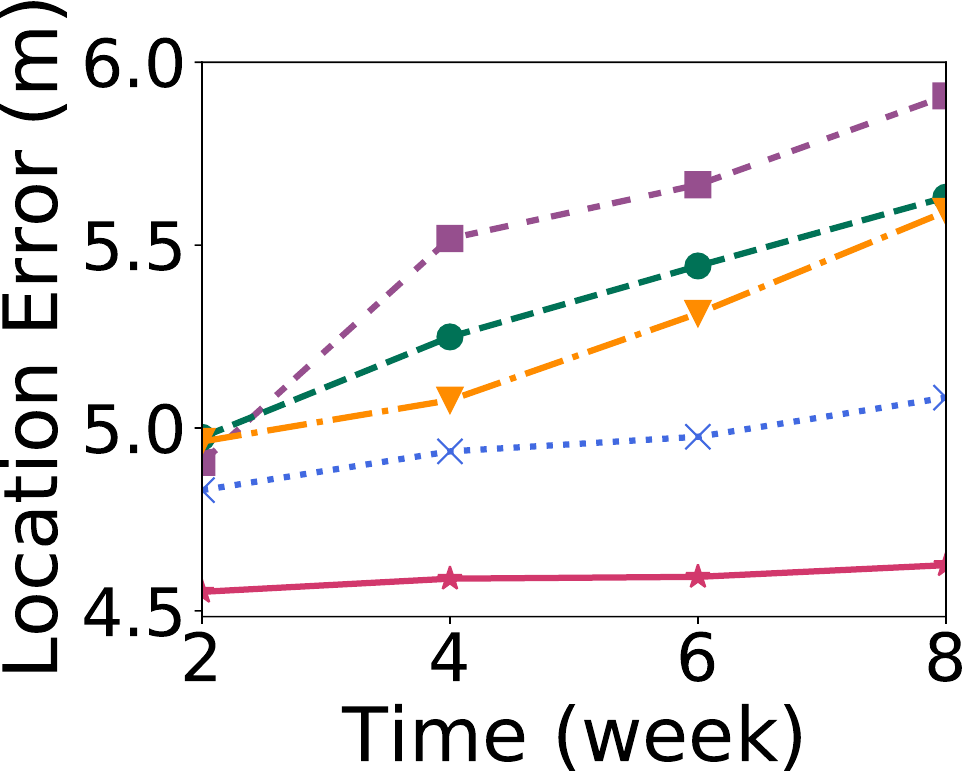}} 
    \hspace{0.02in}
    \subfigure[GF]{\includegraphics[width=0.21\textwidth]{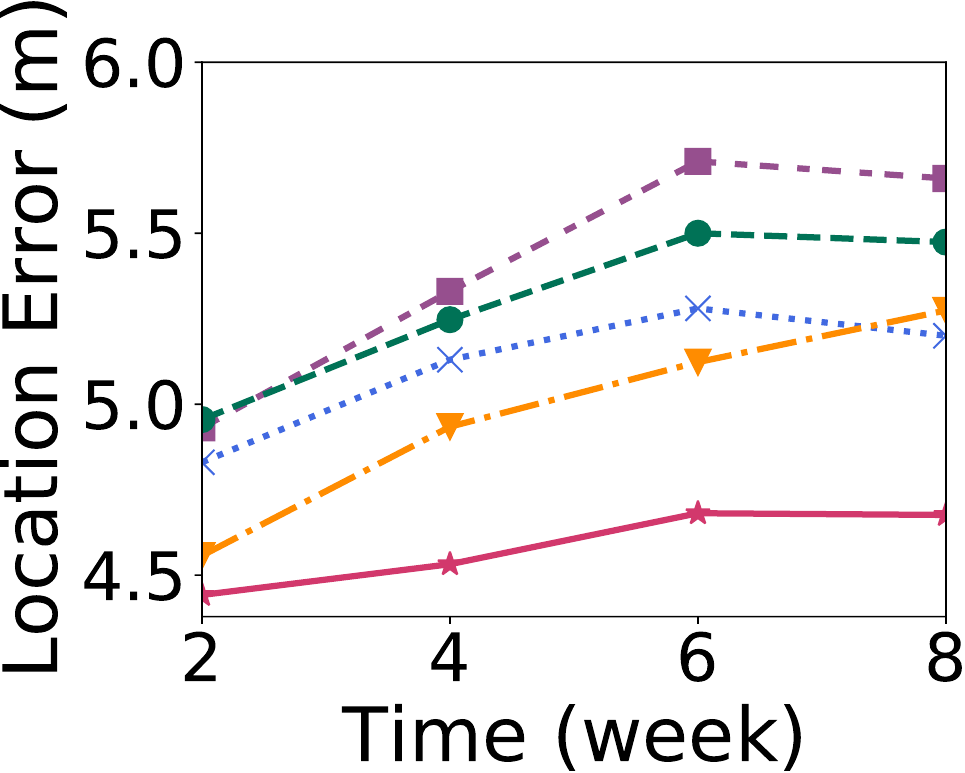}} 
    \hspace{0.02in}
    \subfigure[1F]{\includegraphics[width=0.21\textwidth]{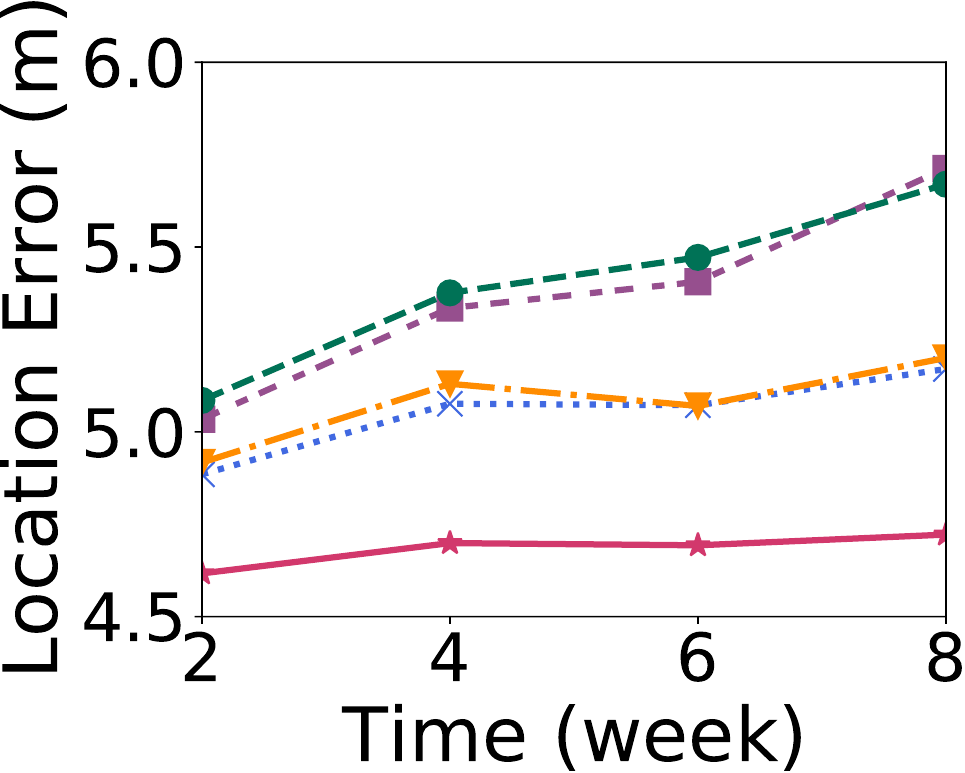}}
    \hspace{0.02in}
    \subfigure[2F]{\includegraphics[width=0.21\textwidth]{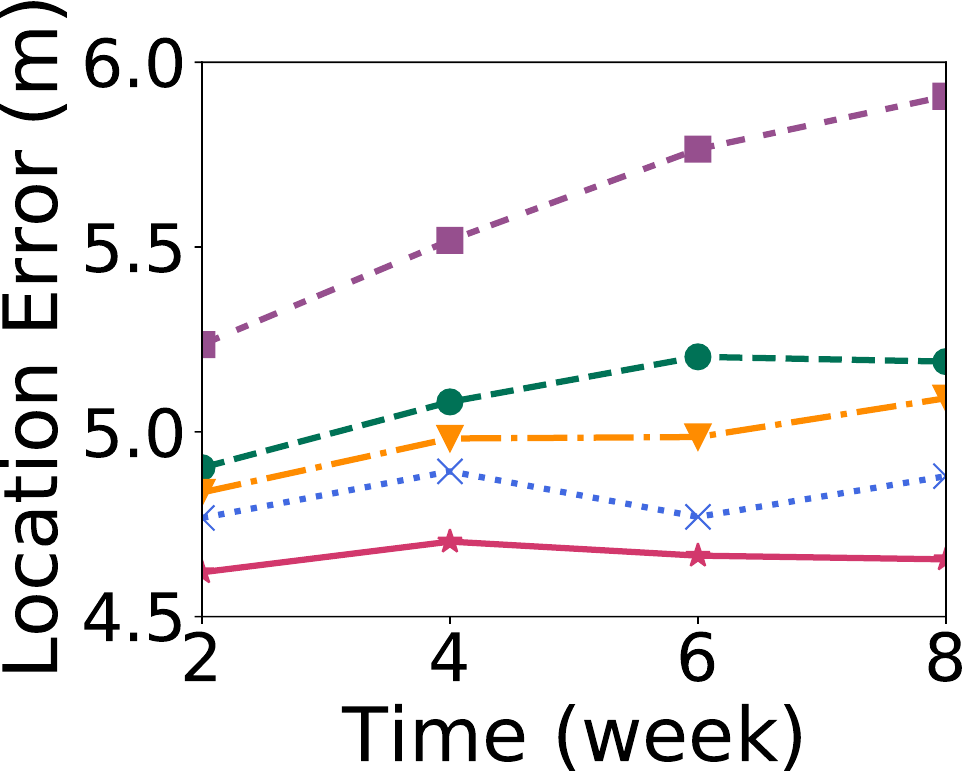}}
    \hspace{0.02in}
    \subfigure
    {\includegraphics[width=0.1\textwidth]{figure/07_eva/2_campus1_legend.pdf}}
    \caption{Location error over eight weeks on four floors in mall A.}
    \label{fig:exp1_mall1}
    \Description{Loc error of mall1, from exp2.}
\end{figure*}

\begin{figure*}
    \centering
    \subfigure[1F]{\includegraphics[width=0.21\textwidth]{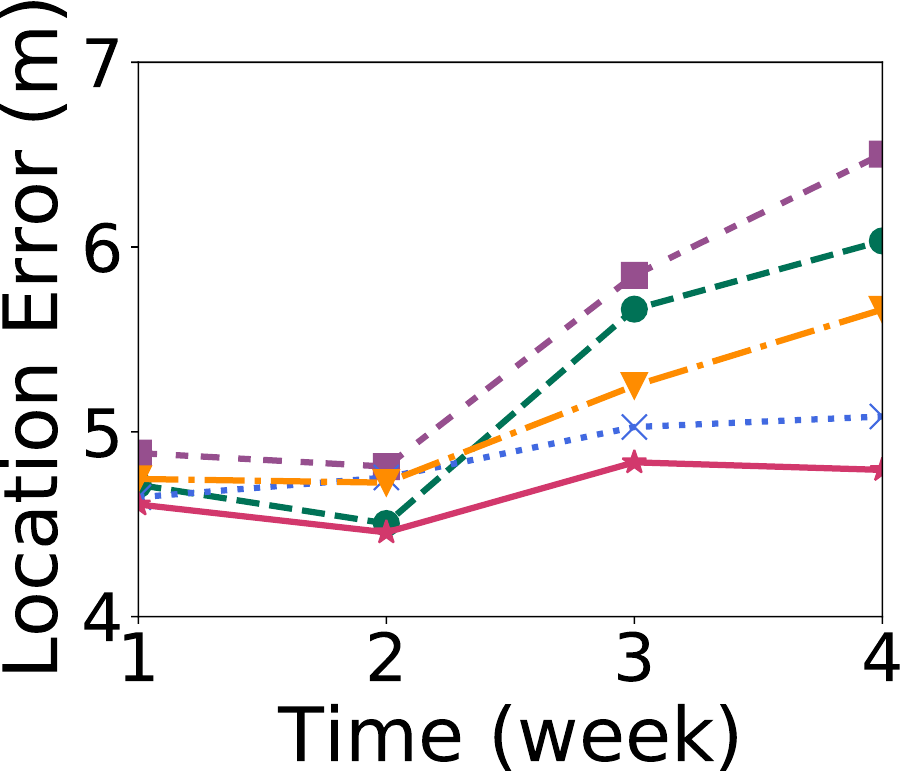}}
    \hspace{0.02in}
    \subfigure[1MF]{\includegraphics[width=0.21\textwidth]{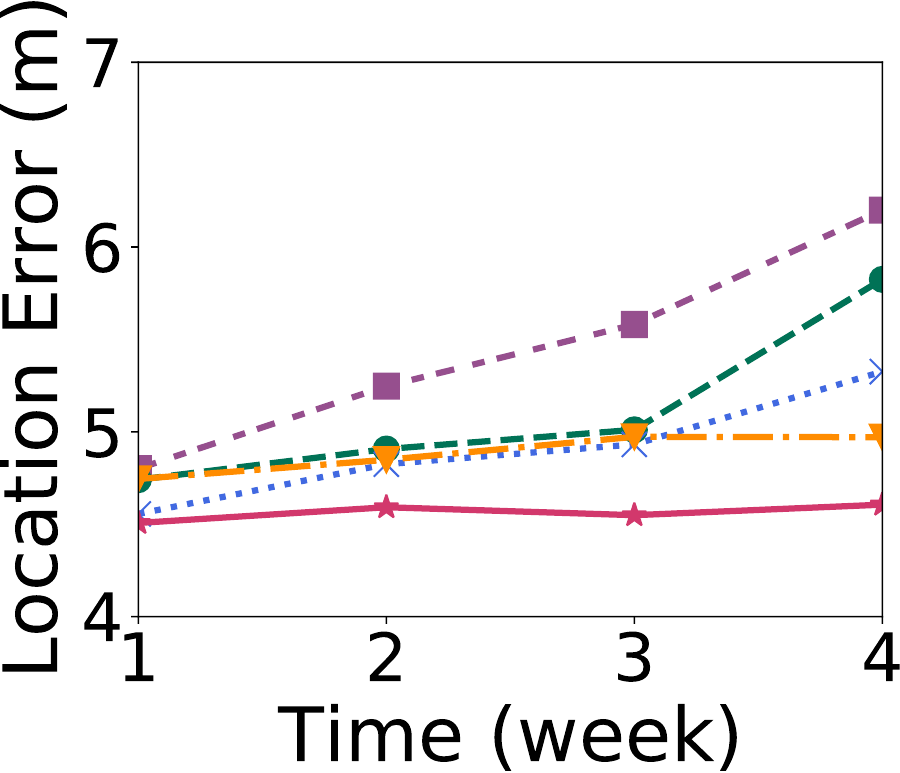}} 
    \hspace{0.02in}
    \subfigure[2F]{\includegraphics[width=0.21\textwidth]{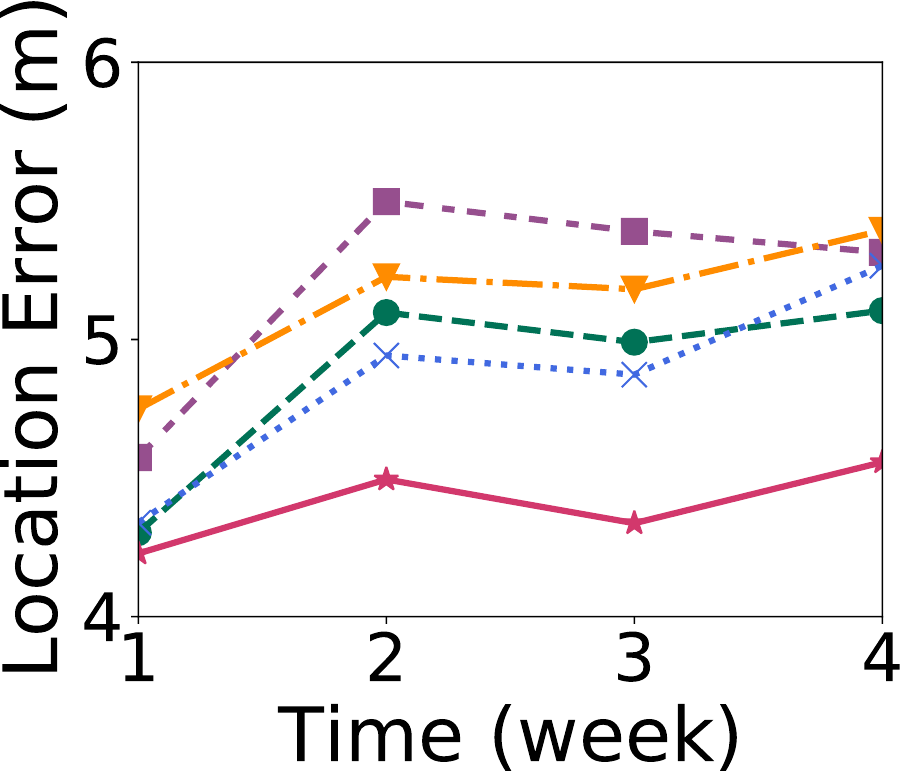}}
    \hspace{0.02in}
    \subfigure[3F]{\includegraphics[width=0.21\textwidth]{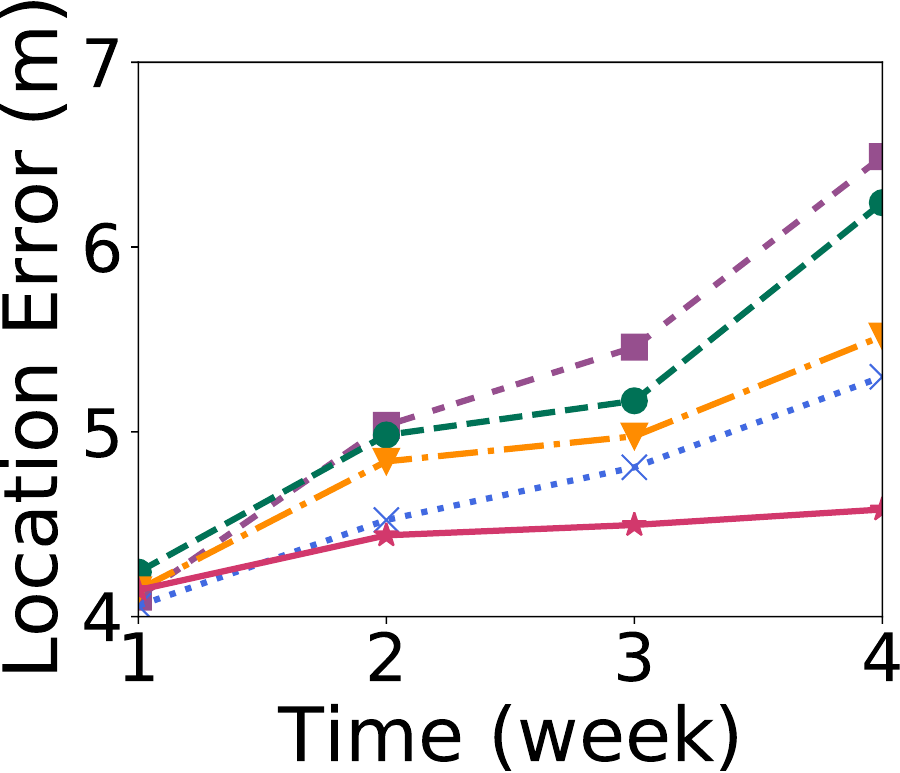}}
    \hspace{0.02in}
    \subfigure
    {\includegraphics[width=0.1\textwidth]{figure/07_eva/2_campus1_legend.pdf}}
    \caption{Location error over eight weeks on four floors in mall B.}
    \label{fig:exp1_mall2}
    \Description{Loc error of mall2, from exp2.}
\end{figure*}

\begin{figure*}
    \centering
    \subfigure[GF]{\includegraphics[width=0.21\textwidth]{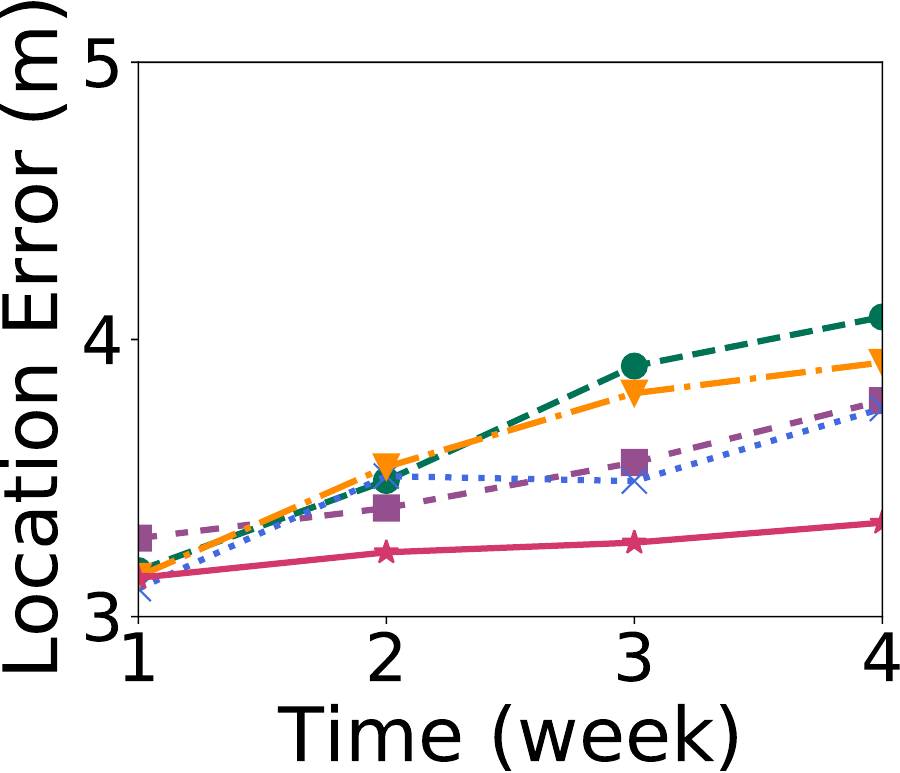}} 
    \hspace{0.02in}
    \subfigure[1F]{\includegraphics[width=0.21\textwidth]{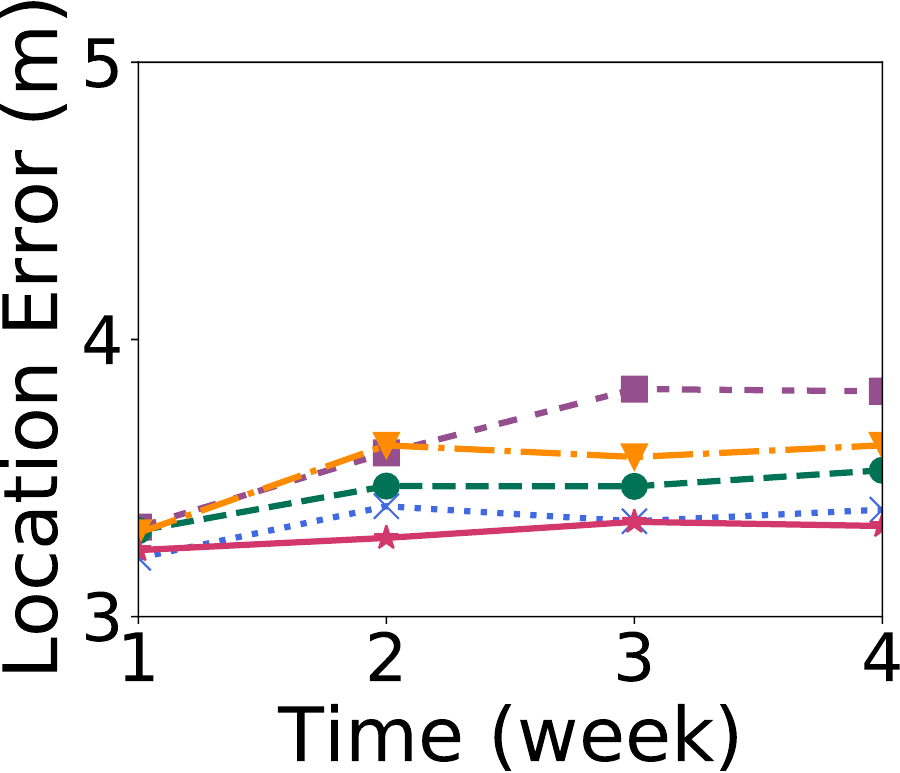}}
    \hspace{0.02in}
    \subfigure[2F]{\includegraphics[width=0.21\textwidth]{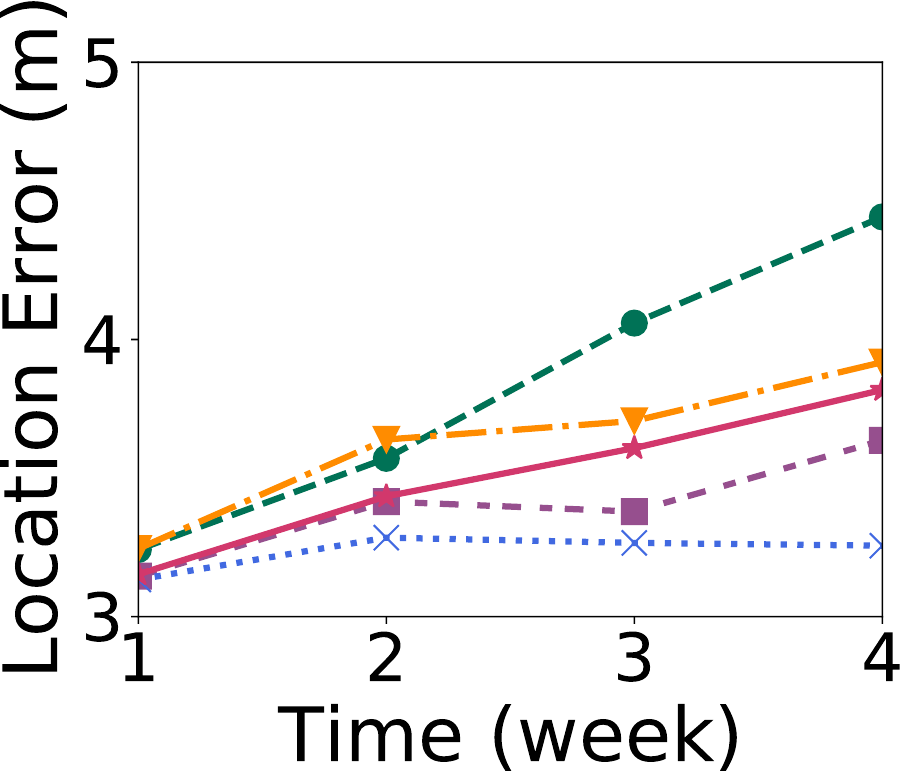}}
    \hspace{0.02in}
    \subfigure[4F]{\includegraphics[width=0.21\textwidth]{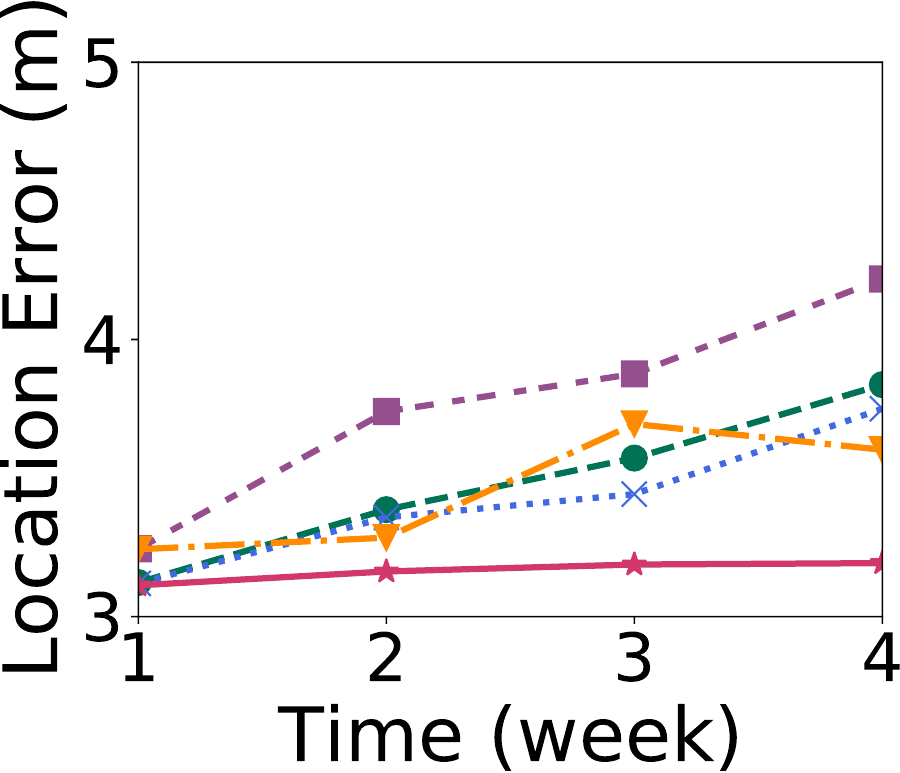}}
    \hspace{0.02in}
    \subfigure
    {\includegraphics[width=0.1\textwidth]{figure/07_eva/2_campus1_legend.pdf}}
    \caption{Location error over eight weeks on four floors in the hospital.}
    \label{fig:exp1_h1}
    \Description{Loc error of h1, from exp2.}
\end{figure*}

\noindent\textbf{Fingerprint updates}: We first evaluate the performance of \n{} and other baseline algorithms for updating \emph{aged} fingerprints, which is measured in the RSS error. As shown in Figure~\ref{fig:exp2}, \n{} outperforms the other algorithms substantially over all four sites.\footnote{A decrease of 3dB in signal strength indicates that the signal power reduces by half.} It indicates the effectiveness of our representation learning for WiFi signal samples and the fingerprint-updating MLP networks built upon the representations. Fidora, however, does not perform well because signal characteristics in neighboring areas can be similar to each other, which is in contrast to the rationale behind its design. The performance of iToLoc, MTDAN, and WiDAGCN is also not satisfactory. While they update RSS values based on the predicted locations of new signals, their location predictions are not as accurate as \n{}. See Figures~\ref{fig:exp1_campus1}--\ref{fig:exp1_h1} for more details on the location prediction accuracy, which shall be explained below. Furthermore, all the baseline algorithms do not consider newly introduced APs in the new signals, thereby negatively affecting the location prediction accuracy and the quality of fingerprint updates.

\begin{table*}
\caption{Summary of the mean (standard deviation) for location prediction errors in different sites.}
\label{tab:loc}
\begin{center}
\resizebox{0.95\textwidth}{!}{
\begin{tabular}{| c | c | c | c | c | c |} 
 \hline
 Site &  \n{} & Fidora & WiDAGCN & MTDAN & iToLoc \\
 \hline
 Campus & \textbf{4.92m} (0.815m) & 5.48m (0.781m) & 5.67m (0.919m) & 6.22m (0.738m) & 6.53m (0.833m) \\ 
 \hline
     Mall A & \textbf{4.44m} (0.974m) & 4.88m (1.010m) & 5.06m (1.071m)& 5.21m (1.268m) & 5.49m (1.541m) \\
 \hline
 Mall B & \textbf{4.39m} (1.252m) & 5.46m (1.872m)& 5.39m (1.997m) & 5.77m (1.825m) & 5.94m (1.808m) \\
 \hline
 Hospital & \textbf{3.06m} (0.603m) & 3.72m (0.466m)& 4.03m (0.704m) & 3.99m (0.840m) & 4.18m (0.597m) \\
 \hline
\end{tabular}
}
\end{center}
\end{table*}

\vspace{1mm}
\noindent\textbf{Location prediction accuracy:} We summarize the average location prediction errors of \n{} and other state-of-the-art algorithms in Table~\ref{tab:loc}. \n{} outperforms other algorithms significantly, with up to 82\% improvement in the location error. This is because \n{} accurately updates fingerprints over time thanks to our novel graph-based representation learning for WiFi signals along with effective link predictions. However, iToLoc and MTDAN do not perform well as their signal features may vary over time, which affects the performance of their time discriminators. Fidora is also not able to correctly classify new signals (or infer their locations) since possibly similar signal characteristics over the neighboring areas may hinder its discriminative power. WiDAGCN does not perform well because its subgraph matching and prediction may be based on outdated fingerprint information. Moreover, they do not include newly added APs in the fingerprints, which affects their performance over time.

\begin{figure*}
    \centering
     \includegraphics[width=0.6\linewidth]{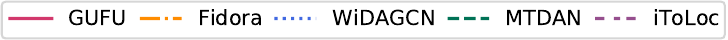}

        \subfigure[Campus]{
        \includegraphics[width=0.23\textwidth]{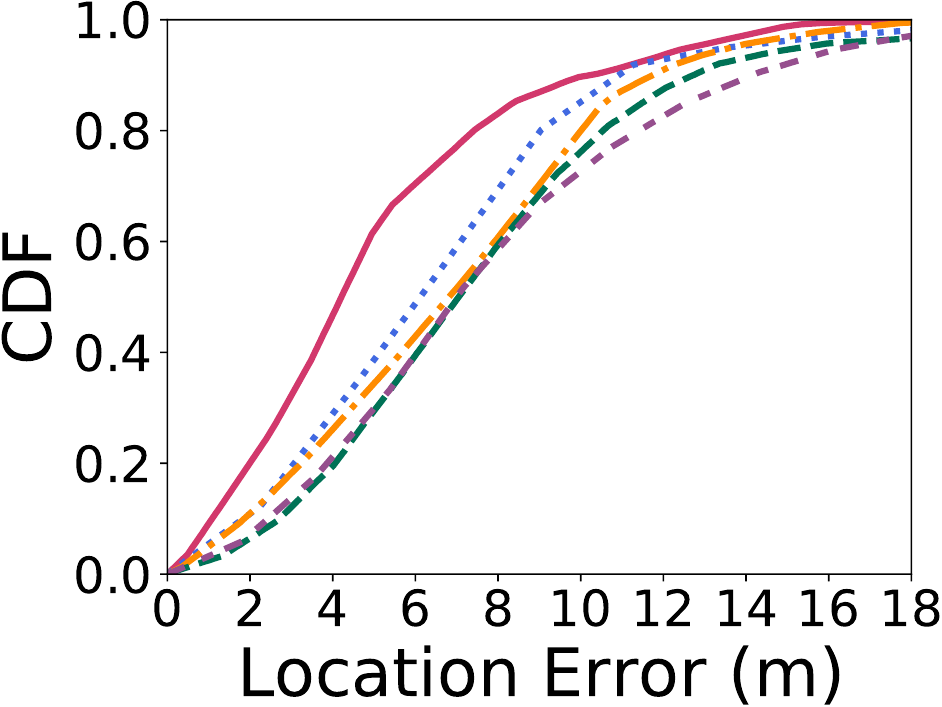}
        }            
        \subfigure[Mall A]{
        \includegraphics[width=0.23\textwidth]{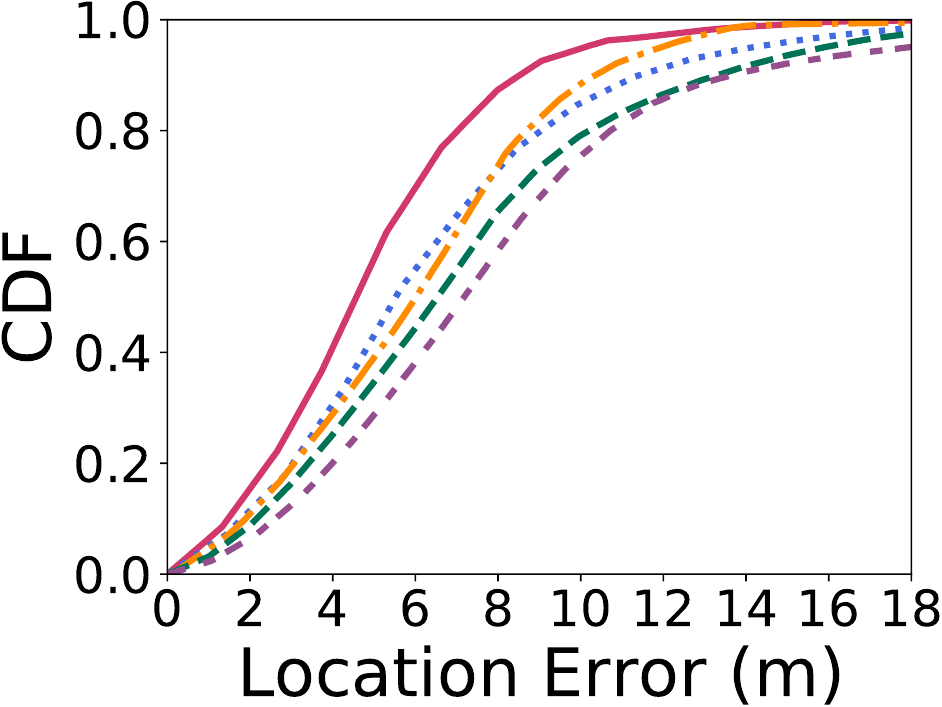}
        }
        \subfigure[Mall B]{
        \includegraphics[width=0.23\textwidth]{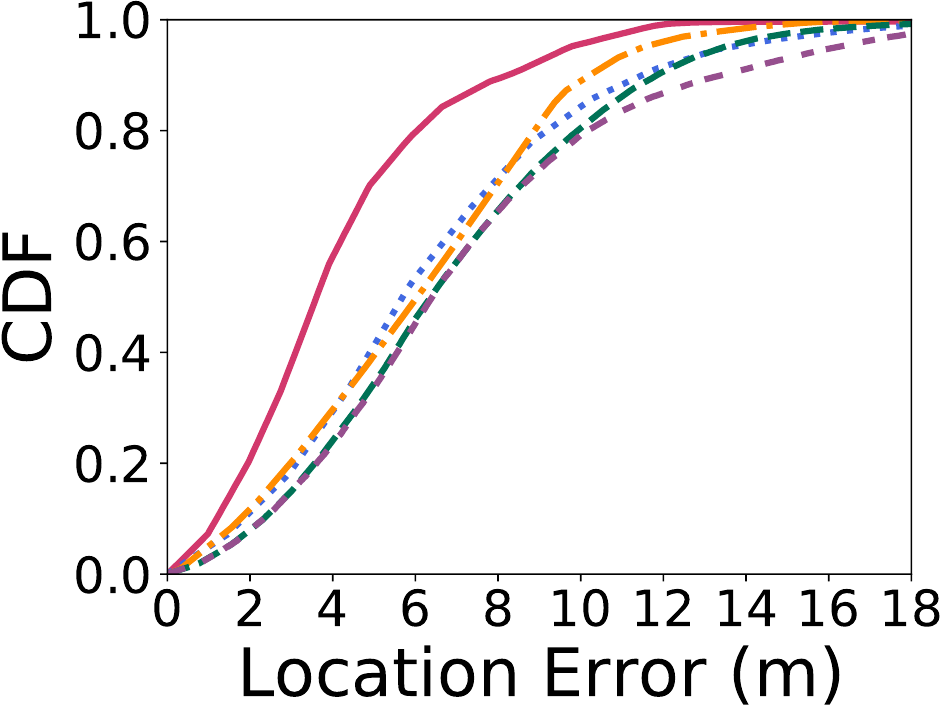}
        }
        \subfigure[Hospital]{
        \includegraphics[width=0.23\textwidth]{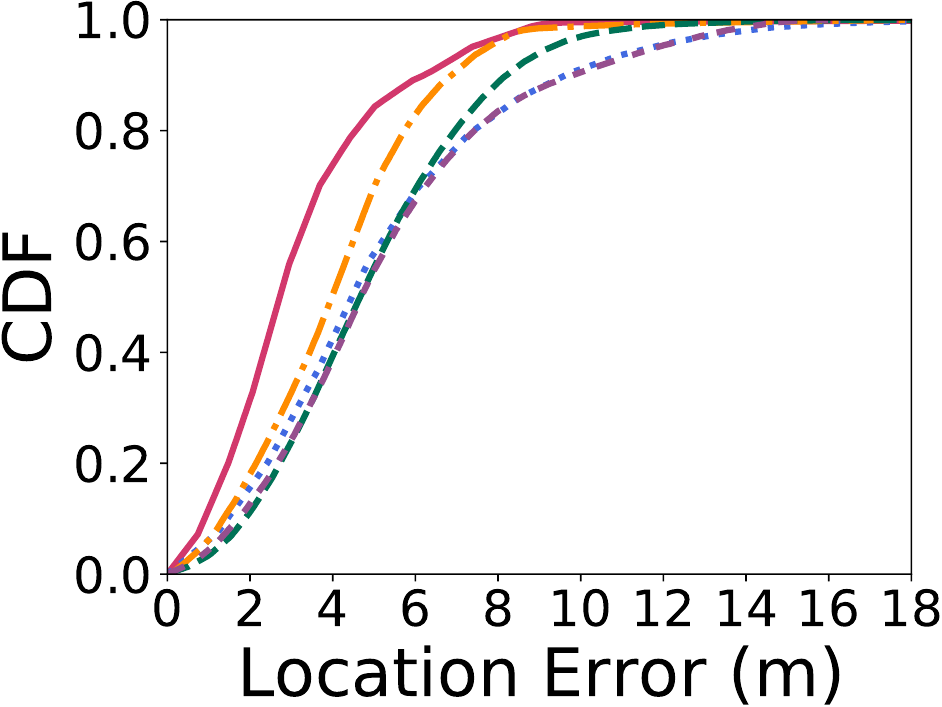}
        }    
    \caption{Comparison of CDF on location error for four different sites}
    \label{fig:cdf_campus1}
    \Description{Four CDFs, from exp3.}
\end{figure*}

We further evaluate the accuracy of location prediction week by week for newly collected signals in all sites and present the results in Figures~\ref{fig:exp1_campus1}--\ref{fig:exp1_h1}. \n{}'s performance varies at different floors as different floors may exhibit different RF signal environments due to diverse factors such as distinct floor designs, different densities of APs, and their different power levels. Nonetheless, \n{} consistently outperforms other algorithms over the \emph{whole} time period. The improvement of \n{} over the other ones becomes more and more significant as time goes by. This indicates that using \n{}, the fingerprints can be valid for a longer period of time, and a recollection of signals for fingerprint database construction can be done less often. In addition, we show the CDF of location prediction errors for four sites in Figure~\ref{fig:cdf_campus1}. While the prediction errors of \n{} are mostly within 6--8m, the others lead to larger errors. For example, the $90$-th percentile accuracy of \n{} outperforms WiDAGCN, Fidora, MTDAN, and iToLoc by up to 31.6\%, 38.7\%, 39.6\%, and 40.4\%, respectively. All these results validate the effectiveness of \n{} in location prediction and demonstrate its superior performance to the other baseline algorithms. 

% This indicates that using \n{}, the degradation time of the fingerprints can be extended, and recollection of signals for fingerprint construction can be done less often.

\subsection{Ablation Study}
\label{subsec:ablation}

\noindent \textbf{Time required for each update:} We recorded the time required for each update and report the results in Table~\ref{tab:time}. The results show that \n{}'s update time for each batch of new data is similar to the existing models like iToLoc and MTDAN, while being significantly faster than Fidora and WiDAGCN.

\begin{table}
\caption{Time needed (in seconds) for each update over eight weeks on the campus.}
\label{tab:time}
\begin{center}
\begin{tabular}{| c | c | c | c | c | c | c | c | c | c |} 
 \hline
Method & Init & Update1 & Update2 & Update3 & Update4 & Update5 & Update6 & Update7 & Update8\\
 \hline
  GUFU & 561.64 & 94.19 & 96.63 & 99.11 & 91.24 & 94.68 & 92.65 & 98.84 & 95.10 \\
 \hline
  Fidora & 692.81 & 102.18 & 109.49 & 115.39 & 106.16 & 108.97 & 102.56 & 102.28 & 99.30 \\
 \hline
  iToLoc & 464.49 & 86.36 & 81.93 & 85.05 & 88.62 & 76.79 & 75.40 & 74.72 & 89.08 \\
 \hline
  MTDAN & 537.78 & 92.28 & 93.32 & 96.83 & 93.64 & 97.43 & 92.46 & 95.22 & 93.29 \\
 \hline
  WiDAGCN & 987.30 & 149.70 & 152.65 & 169.33 & 158.86 & 161.31 & 165.36 & 163.81 & 156.94 \\
 \hline
\end{tabular}
\end{center}
\end{table}

\begin{figure*}
    \centering
    \includegraphics[width=0.8\linewidth]{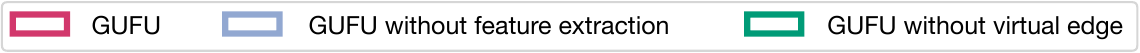}

        \subfigure[]{\includegraphics[width=0.23\textwidth]{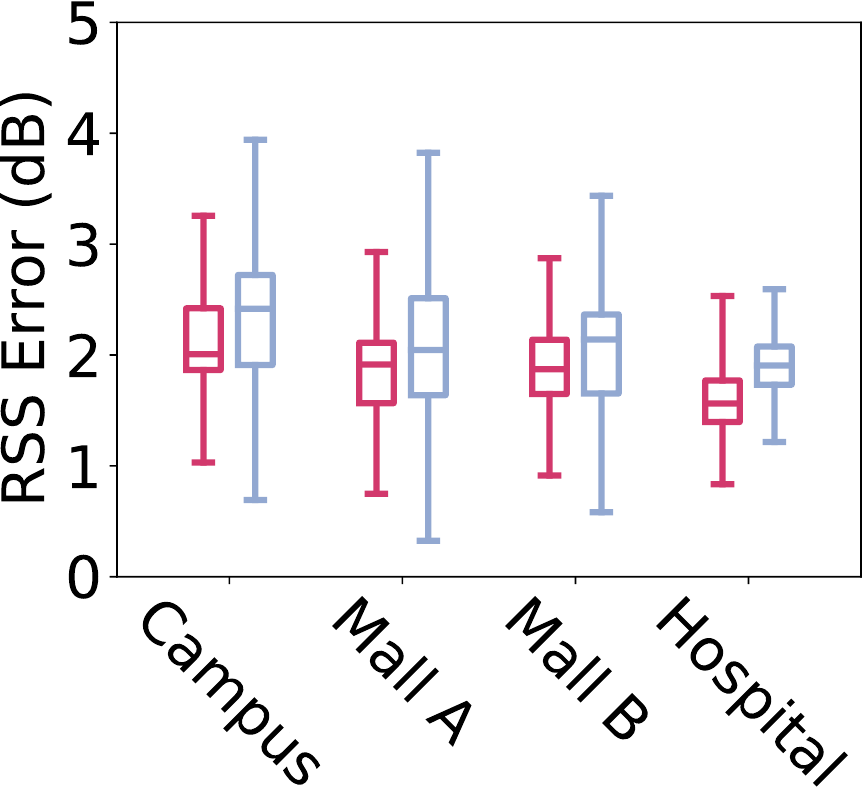}}
        \hspace{0.02in}
        \subfigure[]{\includegraphics[width=0.23\textwidth]{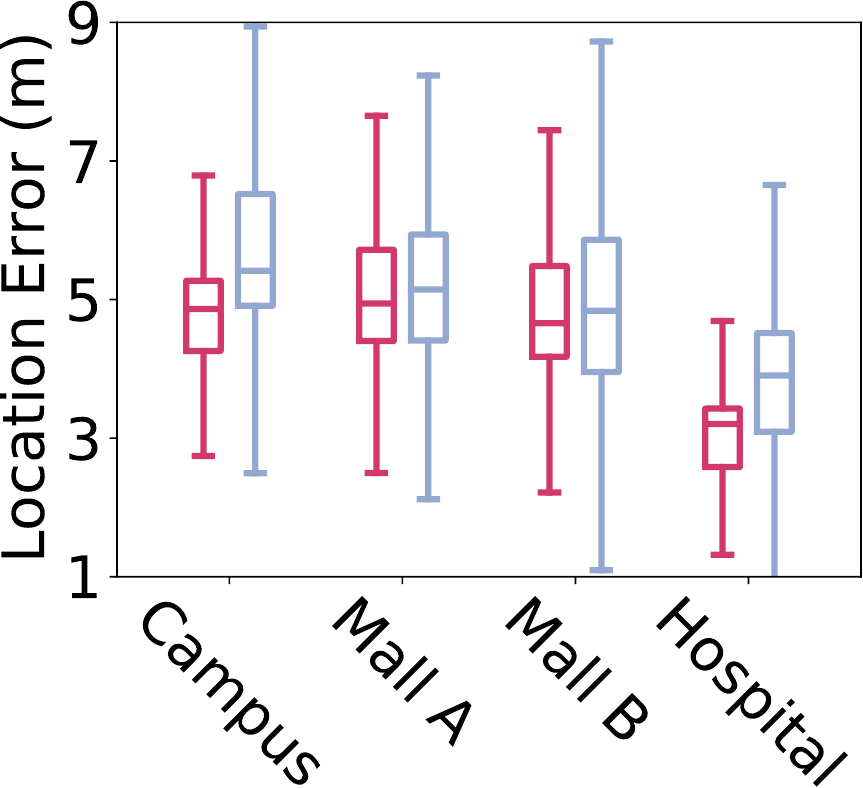}} 
        \hspace{0.02in}
        \subfigure[]{\includegraphics[width=0.23\textwidth]{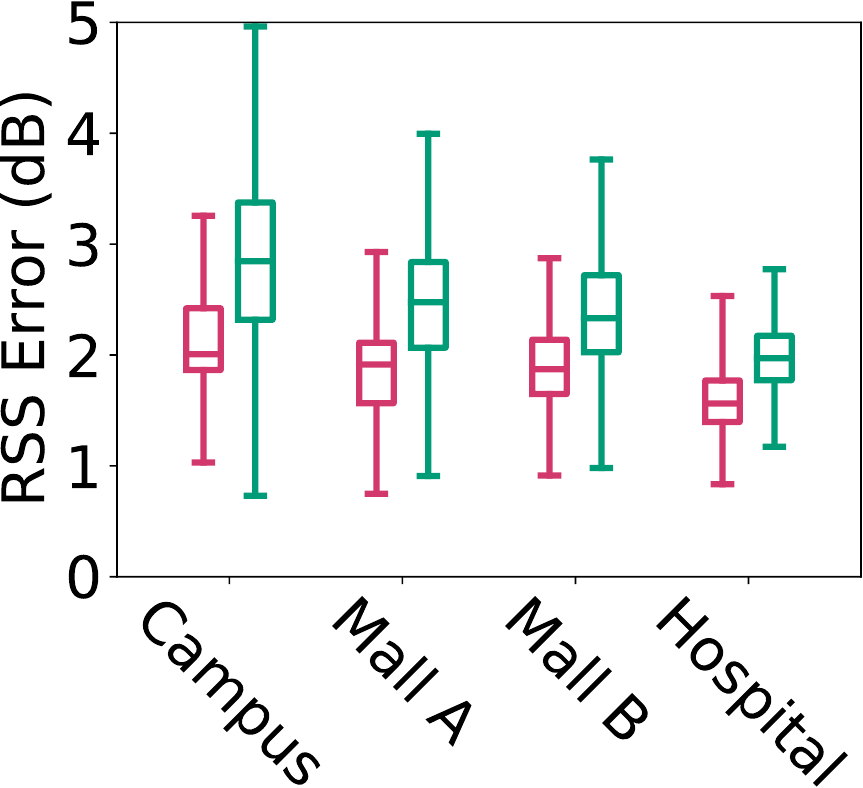}}
        \hspace{0.02in}
        \subfigure[]{\includegraphics[width=0.23\textwidth]{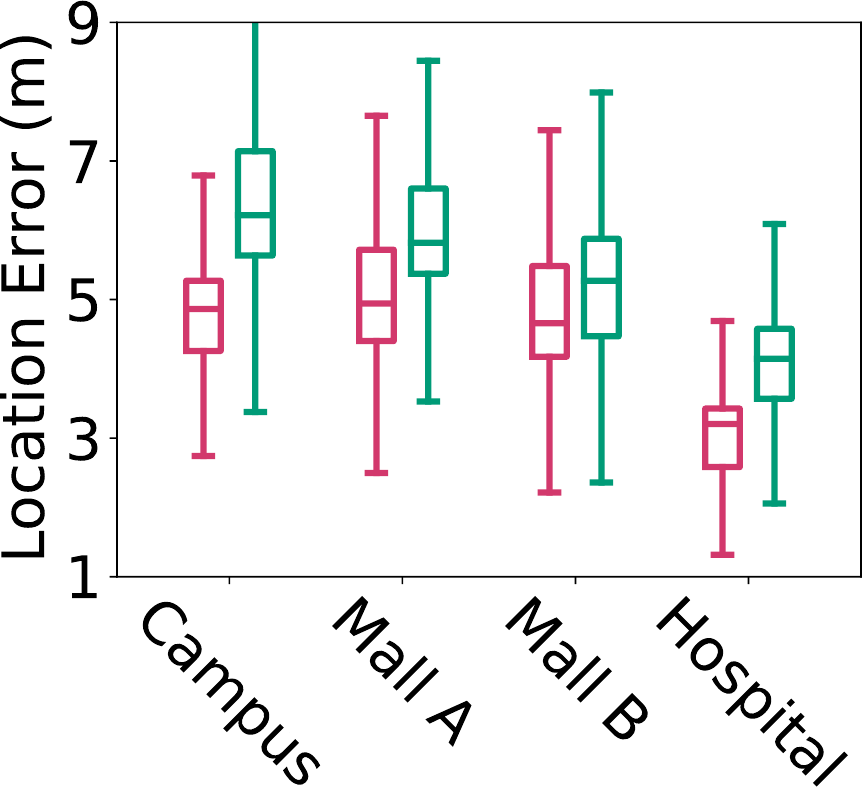}}
        \caption{Ablation study of \n{}. (a) and (b) \n{} (without feature extraction); (c) and (d) \n{} (without virtual edges).}
    \label{fig:ablation_12}
    \Description{Ablation study for feature extractor and virtual edges, from aba1.}
\end{figure*}

\noindent\textbf{Impact of RSS-feature extractor:} We use an autoencoder as the feature extractor to learn the information carried by the fingerprints. The rationale behind our choice of the autoencoder is that it has the capability to ``reconstruct'' signals from the extracted signal features using its decoder. This structure aligns well with our system design, where the extracted signal features are intended to predict the signal strengths needed for fingerprint updates. To show the benefits of using such a feature extractor, we quantitatively compare the location error and RSS error with and without this module in Figure~\ref{fig:ablation_12}(a)--(b). For \n{} without feature extraction, we directly use the normalized RSS values as the node features on the graph, i.e., adding an offset of 120dBm and then dividing the obtained value by 120. As shown in the plots, \n{} achieves a substantial improvement over its version without the feature extractor in RSS error and location error, by up to 26.6\% and 18.9\%, respectively. This demonstrates that the initialization of node features with the feature extractor indeed helps to better capture similarities between WiFi signal samples and thus improves the location prediction and fingerprint updating results.

\begin{figure*}
    \centering
     \includegraphics[width=0.8\linewidth]{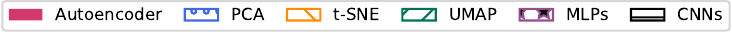}

        \subfigure[Campus]{
        \includegraphics[width=0.22\textwidth]{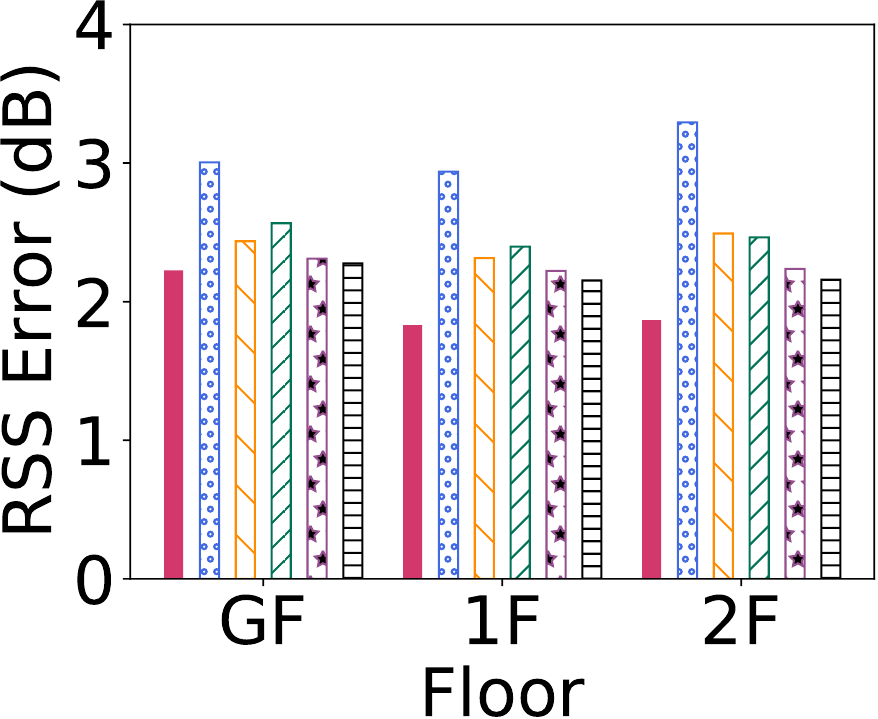}
        }            
        \subfigure[Mall A]{
        \includegraphics[width=0.22\textwidth]{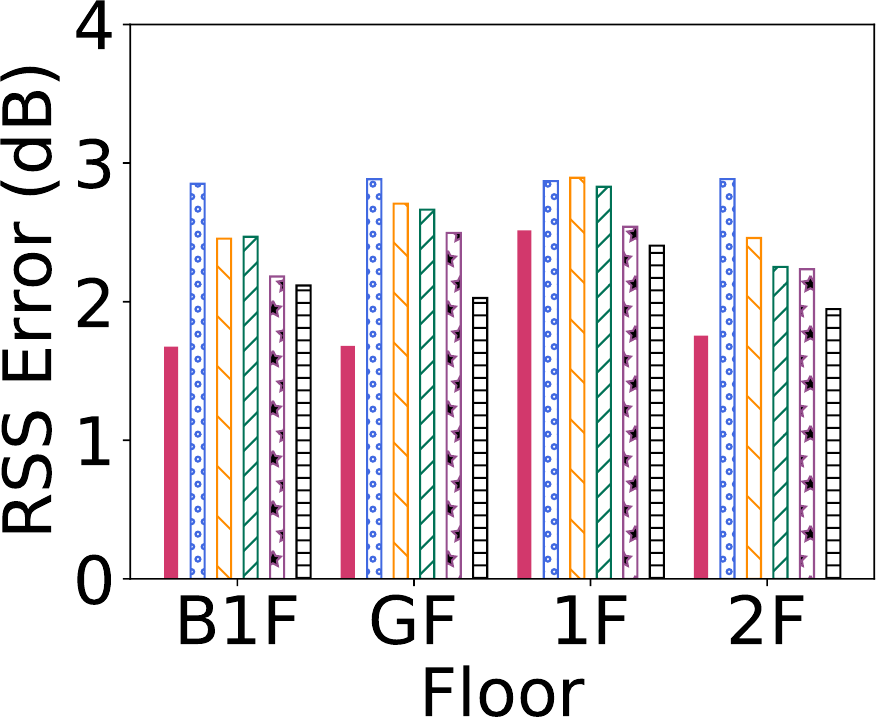}
        }
        \subfigure[Mall B]{
        \includegraphics[width=0.22\textwidth]{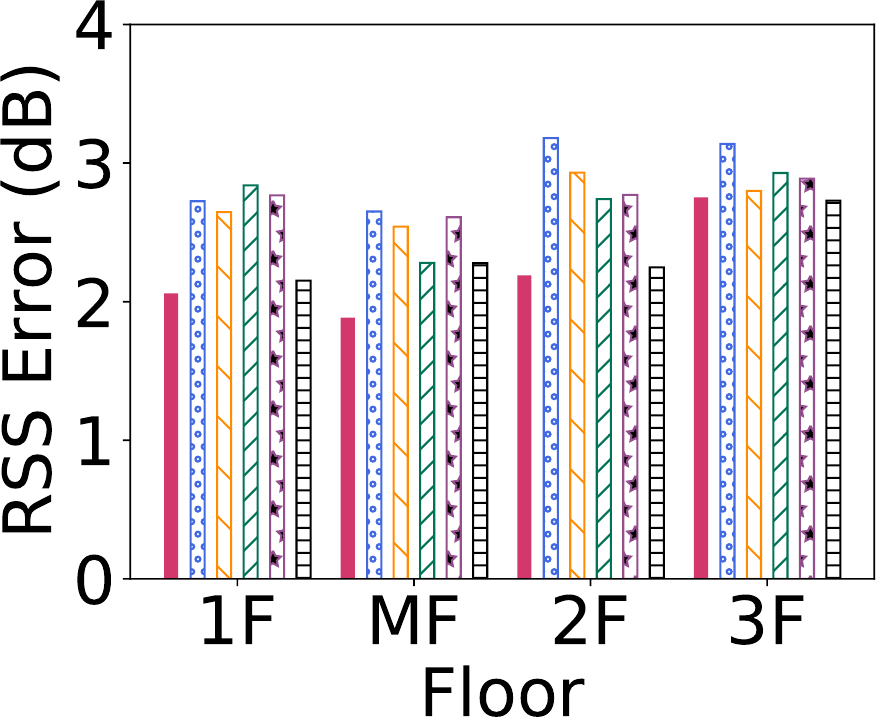}
        }
        \subfigure[Hospital]{
        \includegraphics[width=0.22\textwidth]{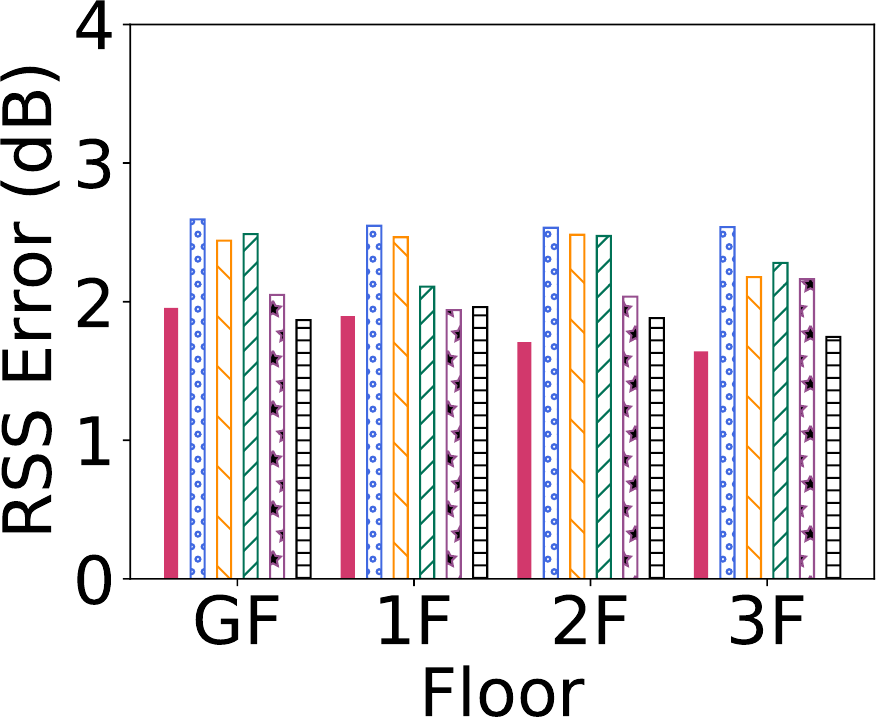}
        }    
    \caption{Summary of RSS error on different floors of the four sites using different feature extractors. }
    \label{fig:exp2_ablation}
    \Description{Ablation study for different feature extractors, from aba2.}
\end{figure*}

We also observe that the autoencoder extracts signal features that \emph{most effectively} represent the original signal strengths, as shown in Figure~\ref{fig:exp2_ablation}. We compare the autoencoder against other popular methods for obtaining fixed-size, lower-dimensional features from a longer list of input signal strengths. Specifically, we consider statistical methods such as principal component analysis (PCA), t-distributed stochastic neighbor embedding (t-SNE), and uniform manifold approximation and projection (UMAP), and learning-based ones such as multilayer perceptrons (MLPs) and convolutional neural networks (CNNs). Here we do not consider more complex models such as generative adversarial networks and vision transformers, because our model needs to be retrained to adapt to changes in the dimensions of input signal strengths resulting from modifications in APs. As shown in Figure~\ref{fig:exp2_ablation}, \n{} having the autoencoder as the feature extractor achieves the \emph{best} performance.

\begin{table}
\caption{Average training time (in seconds) for \n{} to converge using different initialization schemes.}
\label{tab:ablation_training_time}
\begin{center}
\begin{tabular}{| c | c | c | c | c |} 
 \hline
Initialization & Campus & Mall A & Mall B & Hospital\\
 \hline
 Weighted Average & 561.64 & 193.41 & 1205.36 & 104.29 \\ 
 \hline
 Zero Initialization & 813.72 & 269.73 & 1807.44 & 130.16 \\
 \hline
 One Initialization & 801.231 &  295.72 & 1755.39 & 128.51 \\
 \hline
 Random Initialization & 603.20 & 199.64 & 1373.89 & 105.59 \\
 \hline
\end{tabular}
\end{center}
\end{table}

\begin{figure}
    \centering
     \includegraphics[width=.45\linewidth]{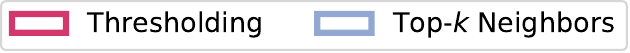}

        \subfigure[RSS Error]{\includegraphics[width=0.3\textwidth]{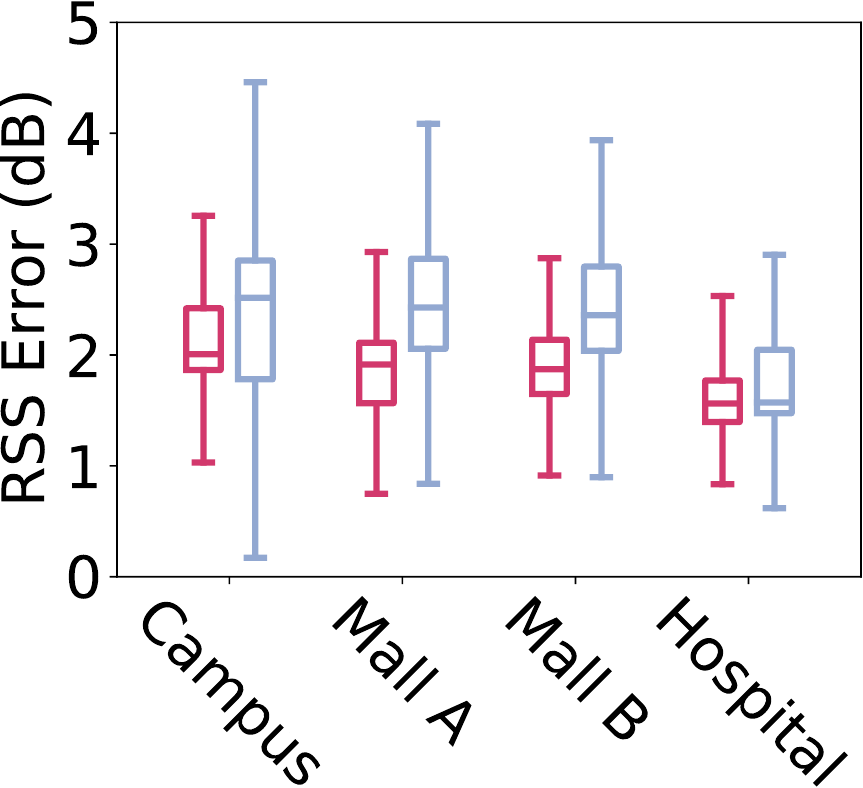}}
        \hspace{0.2in}
        \subfigure[Location Error]{\includegraphics[width=0.31\textwidth]{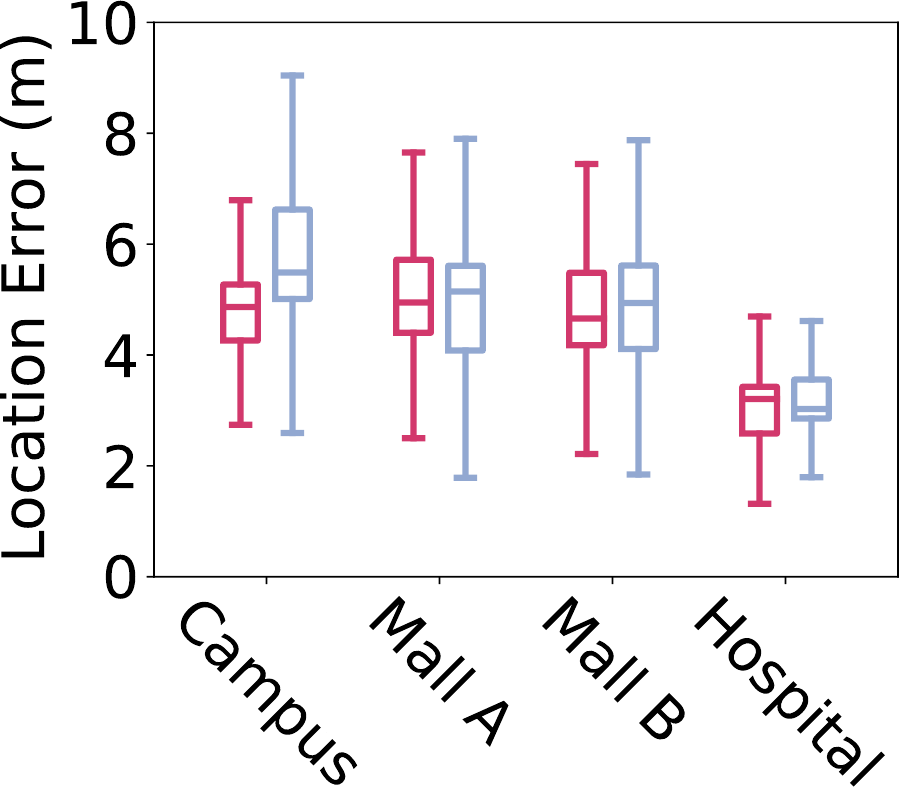}} 
        \caption{Thresholding ($\sigma=0.95$) versus top-$k$ ($k$=10) neighbors in virtual edge creation.}
    \label{fig:ablation_thresh_knn}
    \Description{Ablation study for different thresholding method, from aba3.}
\end{figure}

\vspace{1mm}
\noindent \textbf{Choice of initialization schemes for node features:} For graph initialization, our purpose is to initialize the node features of both sampled nodes and AP nodes in a way that reflects their proximity relationships in physical space. Thus, for the sample nodes, we initialize their feature vectors using the output from the feature extractor, which takes their normalized RSS values from the APs as input. For each AP node, we initialize its feature vector as a weighted average of the feature vectors of its connected sample nodes (i.e., those in its immediate neighborhood), with weights being proportional to the edge weights that are (shifted) RSS values from the AP. That is, the feature vector of each AP is initialized as the result of an information aggregation from its connected sample nodes while also reflecting their proximity relationships in physical space (according to the RSS values).

We have also examined the impact of different initializations on the training time of our GNN model. Specifically, in addition to our weighted averaging, we consider (1) zero-initialization, where each entry is set to 0, (2) one-initialization, where each entry is set to 1, and (3) random initialization, where each entry is assigned a value that is chosen uniformly at random from [0, 1]. As shown in Table~\ref{tab:ablation_training_time}, our weighted averaging achieves the \emph{fastest} training time, which supports the rationale behind our weighted averaging.

\vspace{1mm}
\noindent\textbf{Impact of virtual edges:} The virtual edges enable direct interaction between sample nodes with similar signal features. To validate the effectiveness of this design, we compare \n{}'s performance with and without virtual edges in Figure~\ref{fig:ablation_12}(c)--(d). Including the virtual edges reduces the RSS error and the location error by 11.8\% and 15.2\%, respectively. The results indicate that virtual edges help to propagate useful information between similar sample nodes such that node embeddings can be better learned for location prediction and signal updates.

In addition, our virtual edges are created based on the thresholding of cosine similarity values between signal features of two sample nodes. We compare it with creating virtual edges using top-$k$ ($k$=10) nearest neighbors, and show the performance of the two schemes in Figure~\ref{fig:ablation_thresh_knn}. Our thresholding-based method outperforms top-$k$ nearest neighbors substantially, by 16.2\% in RSS error and 22.1\% in location error, respectively. This is because, in top-$k$ nearest neighbors, there may be sample nodes with low feature similarities selected, as illustrated in Figure~\ref{fig:top_k}, which degrades the performance of location prediction and signal updates. With thresholding, the high feature similarity between sample nodes with virtual edges can be guaranteed, leading to high accuracy in location prediction and signal updates.

\begin{figure*}
    \centering
        \subfigure[Campus]{
        \includegraphics[width=0.23\textwidth]{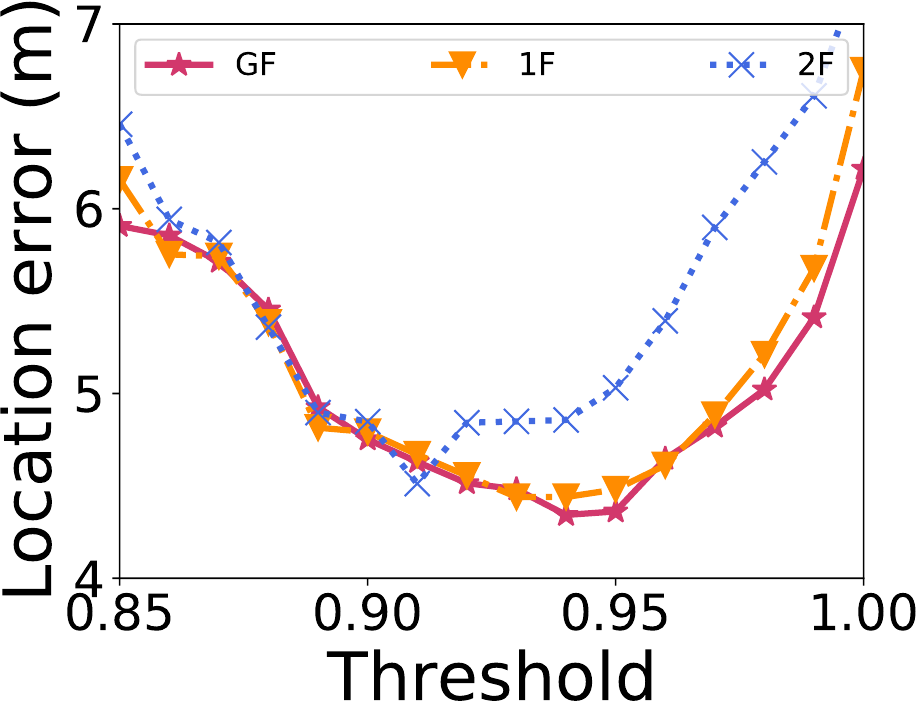}
        }            
        \subfigure[Mall A]{
        \includegraphics[width=0.23\textwidth]{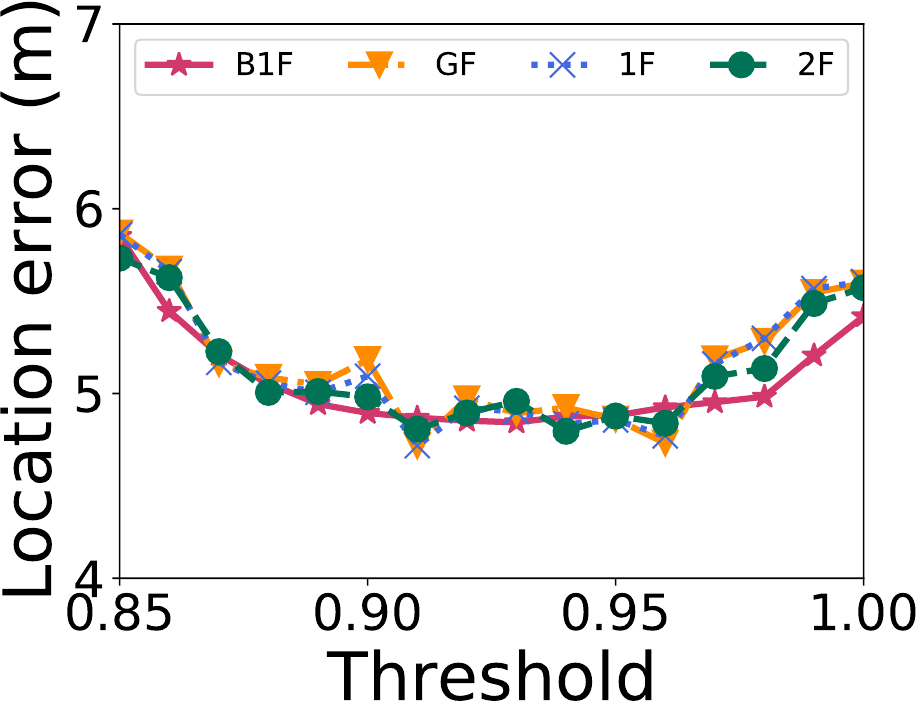}
        }
        \subfigure[Mall B]{
        \includegraphics[width=0.23\textwidth]{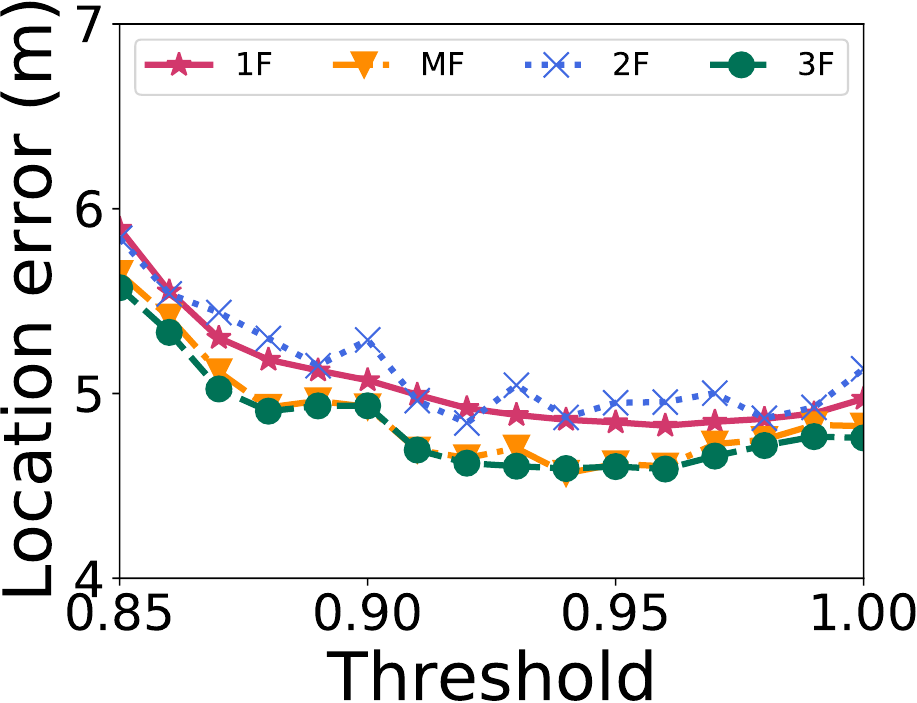}
        }
        \subfigure[Hospital]{
        \includegraphics[width=0.23\textwidth]{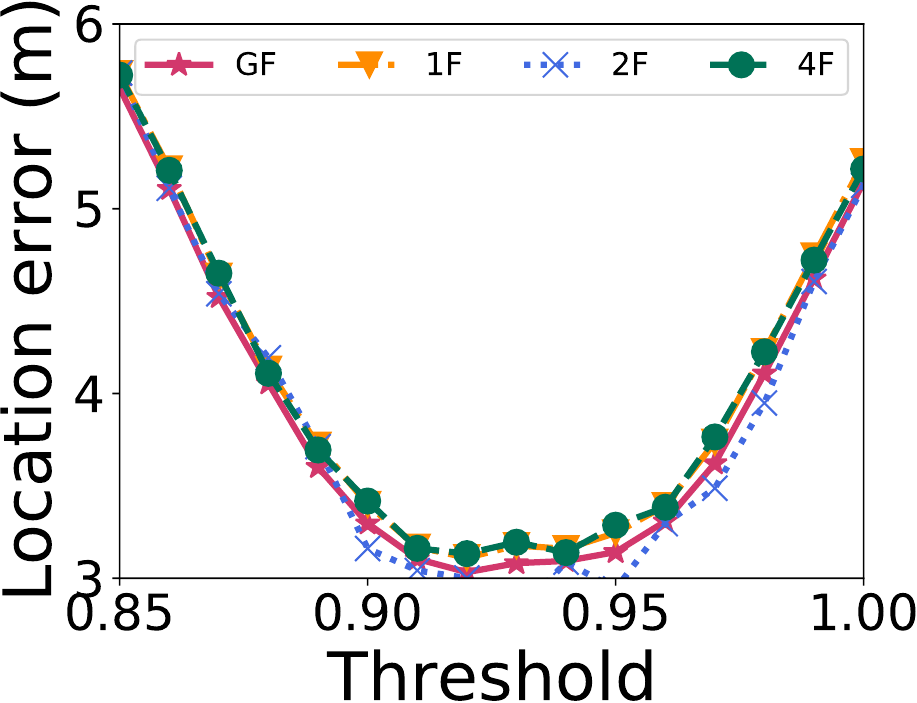}
        }    
    \caption{Location errors on different floors of the four sites using different threshold values of $\sigma$ in virtual edge creation.}
    \label{fig:ablation7}
    \Description{Ablation study for different threshold value for virtual edge creation, from aba4.}
\end{figure*}

In \n{}, the value of threshold $\sigma$ was determined by a hyper-parameter tuning process. We examined the performance of \n{} with varying threshold values and now present the results in Figure~\ref{fig:ablation7}. We observed that threshold values between $0.90$ and $0.97$ yield good performance for \n{}'s location prediction. For simplicity, we set $\sigma = 0.95$ for all sites.

\begin{figure*}
    \centering
    \subfigure[Campus]{\includegraphics[width=0.24\textwidth]{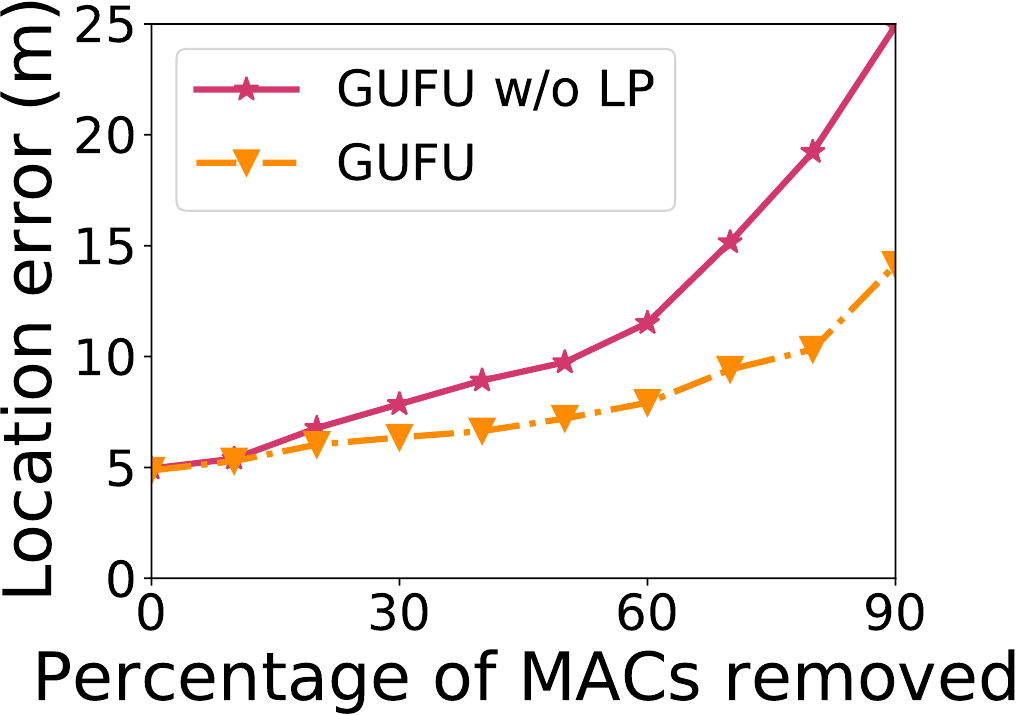}}
    \subfigure[Mall A]{\includegraphics[width=0.24\textwidth]{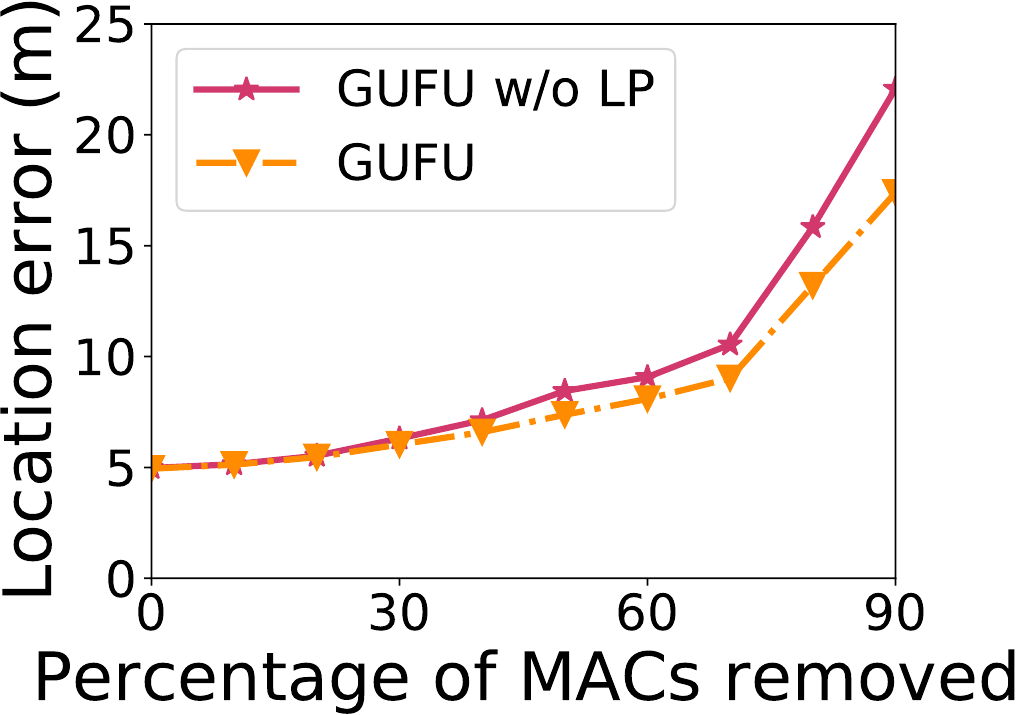}} 
    \subfigure[Mall B]{\includegraphics[width=0.24\textwidth]{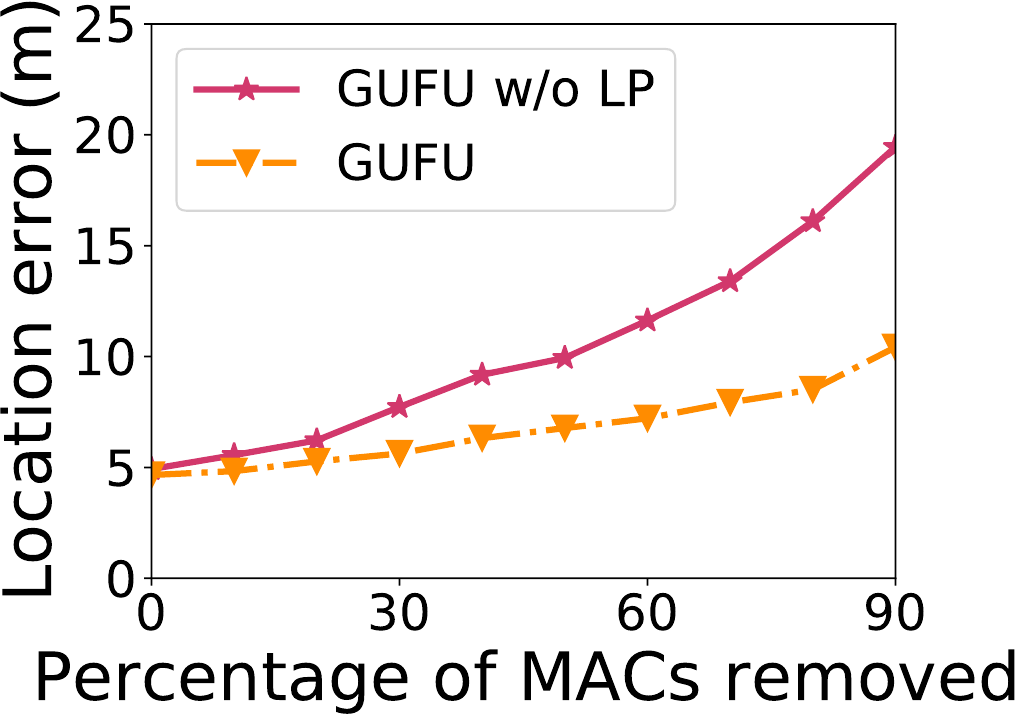}}
    \subfigure[Hospital]{\includegraphics[width=0.24\textwidth]{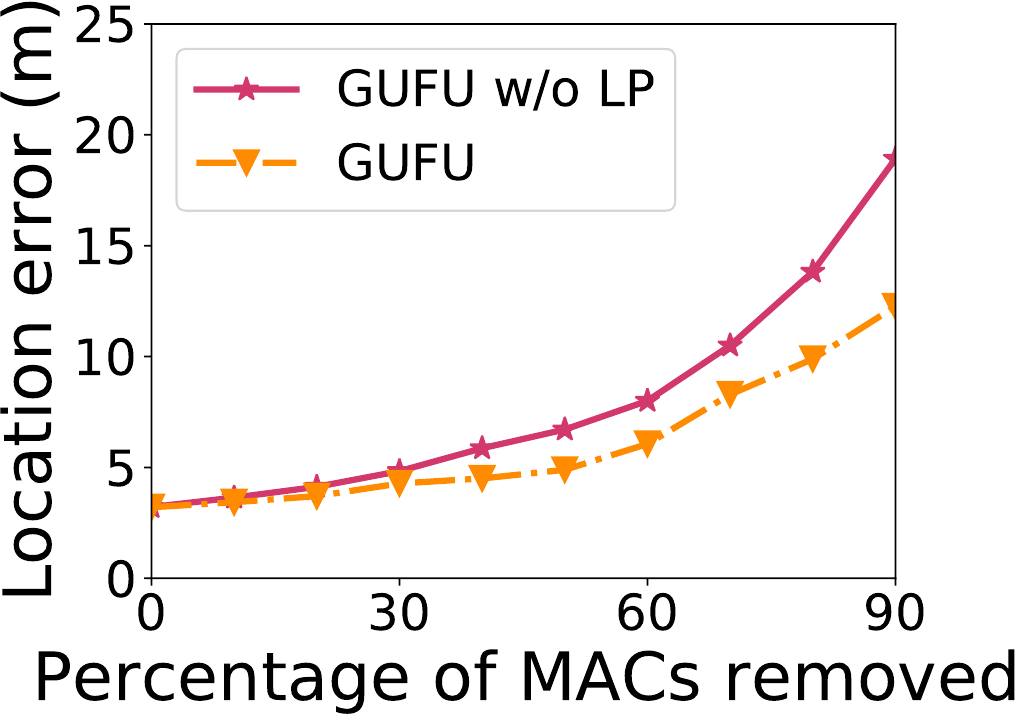}}
    \caption{Location error versus the ratio of MACs removed in four sites.}
    \label{fig:aba3}
    \Description{Ablation study for removing MACs, from aba5.}
\end{figure*}

\vspace{1mm}
\noindent\textbf{Impact of link prediction:} Link prediction enriches the fingerprint database with new connections between AP nodes and sample nodes. To show how much it improves \n{}'s performance, we intentionally remove AP nodes at random in the original fingerprint database. As a result, these MACs will become the newly detected ones in the updating phase. We compare the performance of \n{} and \n{} without link prediction (\n{} w/o LP) in Figure~\ref{fig:aba3}. With more MACs removed from the original fingerprint database, the performance of both schemes degrades. However, compared to \n{} without link prediction, the errors of \n{} increase slowly. This is because \n{} can retain more AP information over time with the link prediction capability. In particular, with up to 90\% of randomly selected MAC addresses removed, \n{} still has an average location error of 15.36m, which is 46.2\% better than that of \n{} without link prediction.

\begin{figure*}
    \centering
     \includegraphics[width=0.5\linewidth]{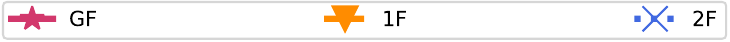}

        \subfigure[]{
        \includegraphics[width=0.22\textwidth]{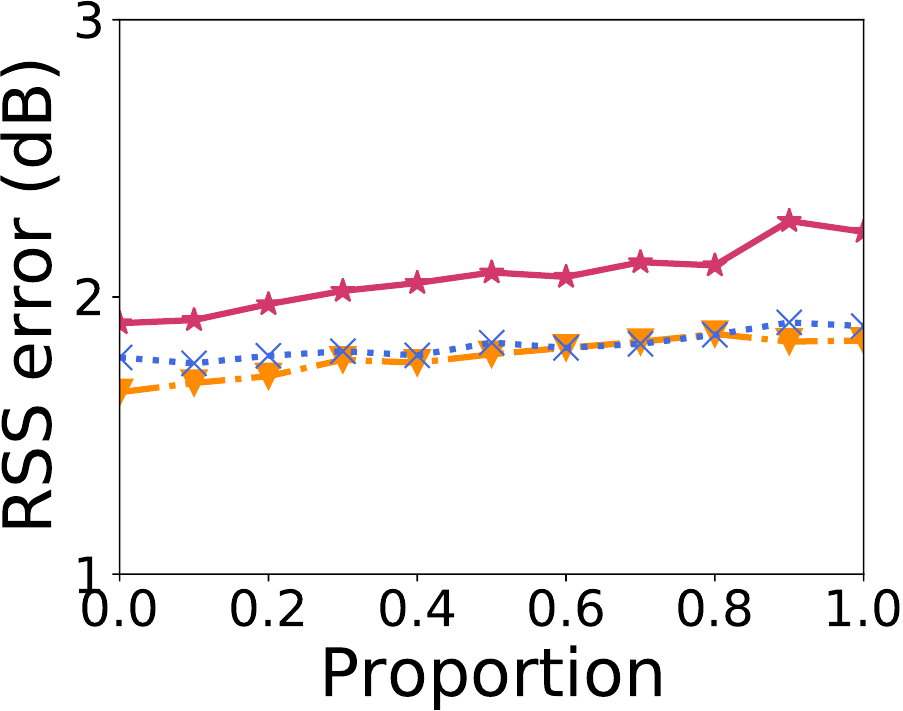}
        }            
        \subfigure[]{
        \includegraphics[width=0.22\textwidth]{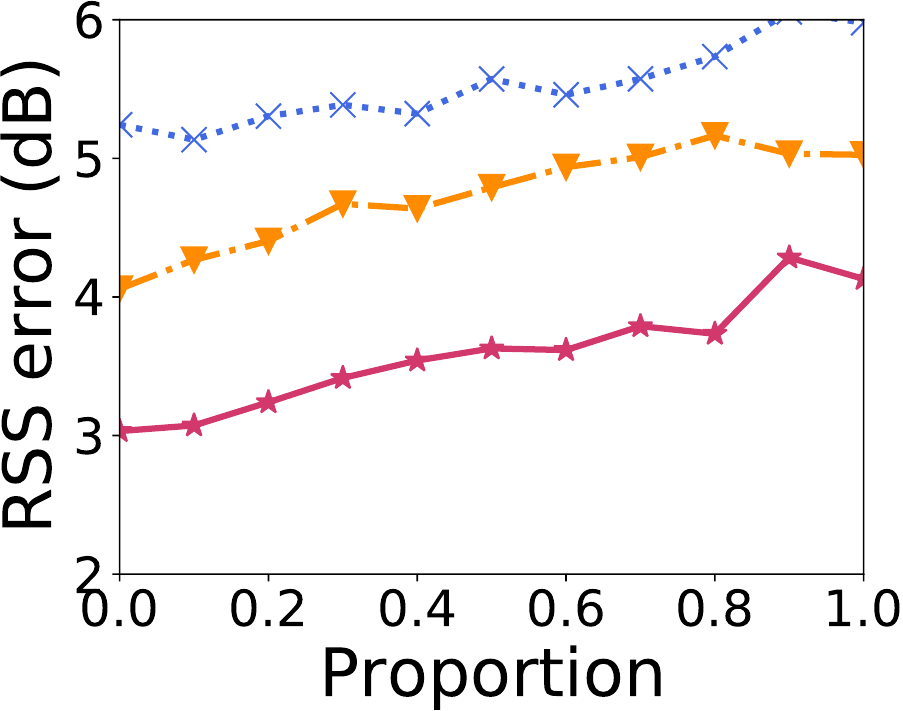}
        }
        \subfigure[]{
        \includegraphics[width=0.22\textwidth]{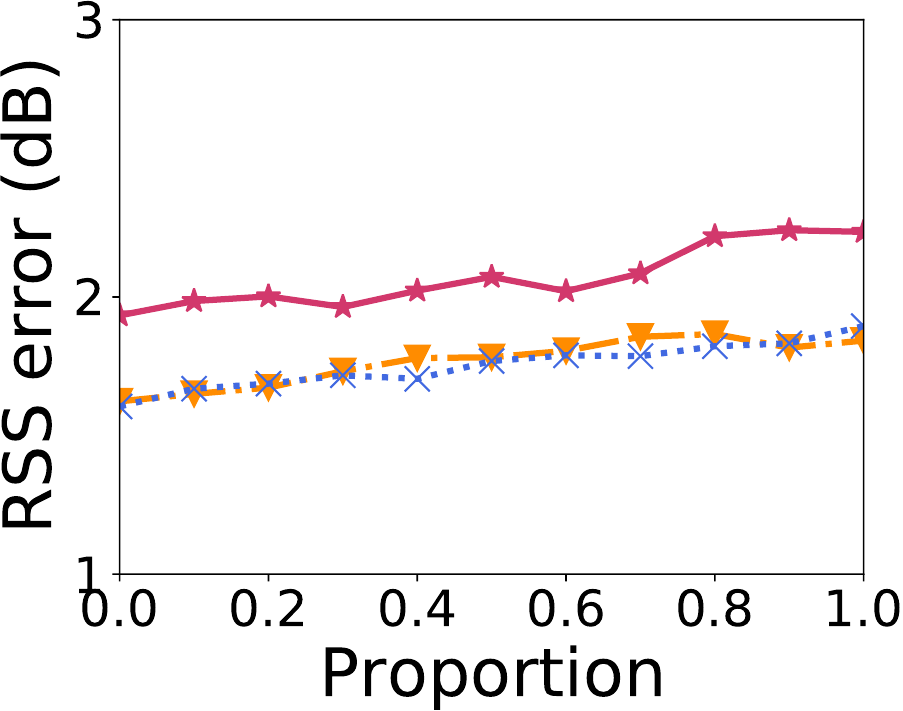}
        }
        \subfigure[]{
        \includegraphics[width=0.22\textwidth]{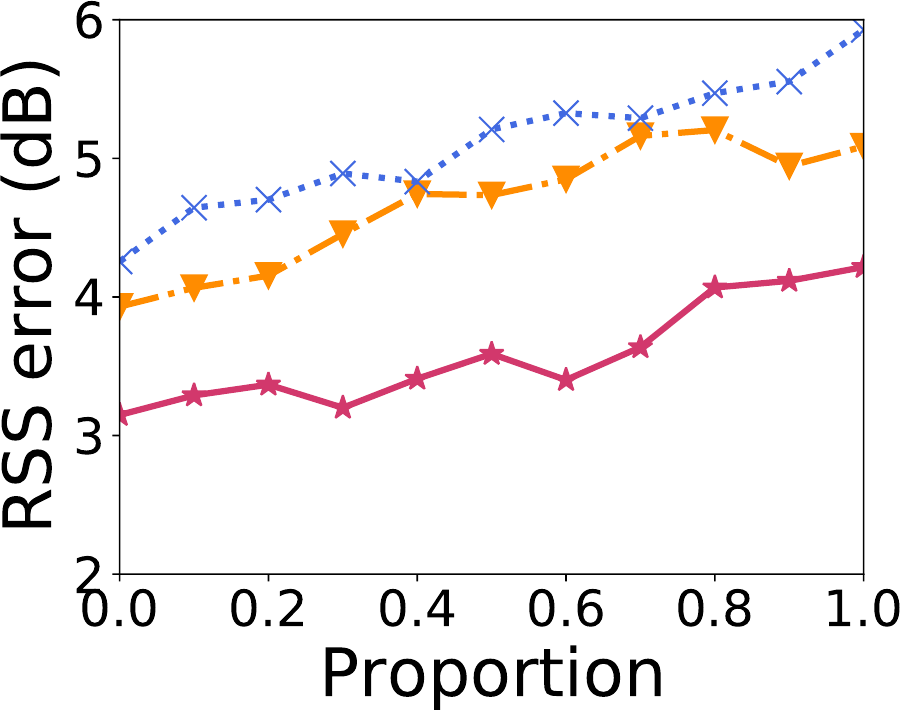}
        }    
    \caption{\n{}'s performance over different AP settings. (a) and (b) new APs. (c) and (d) removed APs.}
    \label{fig:ablation8_proportion}
    \Description{Ablation study for different AP settings, from aba6.}
\end{figure*} 

\vspace{1mm}
\noindent \textbf{Impact of the proportion of evolved APs:} 
\n{} is designed to handle both the addition and removal of APs. To further demonstrate the model's robustness against varying proportions of new and removed APs, we sampled these APs from the campus site, which experienced the most significant AP changes among the four sites in our experiment. In Figure~\ref{fig:ablation8_proportion}, we show \n{}'s performance with varying proportions of added and removed APs. Figure~\ref{fig:ablation8_proportion}(a) and Figure~\ref{fig:ablation8_proportion}(b) show \n{}'s performance considering all removed APs alongside different proportions of new APs, while Figure~\ref{fig:ablation8_proportion}(c) and Figure~\ref{fig:ablation8_proportion}(d) depict the model's performance with all new APs and varying proportions of removed APs. The results indicate that \n{}'s performance remains relatively consistent across different proportions of evolved APs.

It is noteworthy that while \n{}'s performance remains relatively stable across varying proportions of new and removed APs, it tends to perform better with smaller proportions of evolved APs. Consequently, we anticipate that \n{} may reach a limit where the signal prediction error exceeds a certain threshold, necessitating periodic signal recollections. Nonetheless, as shown in Section~\ref{subsec:eva}, the fingerprint database updated using \n{} deteriorates more slowly than those maintained by other state-of-the-art methods. As a result, the time interval between consecutive collections can be longer compared to fingerprints managed by those methods, potentially extending beyond several years. We leave the determination of the exact time interval needed between these consecutive recollections as a future work.

\section{CONCLUSION}
\label{sec:conclude}
In this paper, we propose \n{}, an effective graph-based approach for crowdsourced WiFi fingerprint updates using unlabeled WiFi signals. Our approach relies solely on RSS values from ambient access points (APs) on the site, without requiring any additional information. When a batch of newly collected, yet unlabeled WiFi signals becomes available, \n{} leverages this data to update the existing fingerprints. It is designed to adapt to changes in RSS values, the presence of new APs, and the potential removal of APs in the environment over time. To validate the effectiveness of \n{}, we developed a prototype and conducted extensive evaluations across four different sites over a significant period, ranging from one month to eight months. The experimental results demonstrate that \n{} surpasses other state-of-the-art algorithms in both location prediction and fingerprint updates. Overall, our findings showcase the potential of \n{} in enabling the long-term deployment of a fingerprint database and its automatic enhancement over time, leading to improved fingerprinting-based services.

\begin{acks}
\label{sec:acknowledgement}
This work was supported, in part, by Research Grants Council Collaborative Research Fund (under grant number C1045-23GF). The work of Weipeng Zhuo was supported, in part, by the Guangdong Provincial Key Laboratory of IRADS (2022B1212010006) and UICR0700100-24. The work of Chul-Ho Lee was partially supported by the National Science Foundation under Grant No. 2209921.
\end{acks}

\bibliography{ref}
\bibliographystyle{ACM-Reference-Format}

\end{document}